\documentclass[bibyear]{aa} 

\usepackage{graphicx}
\usepackage{txfonts}
\usepackage{lscape}
\usepackage{booktabs}			

\begin{document}

   \title{Revising the ages of planet-hosting stars}

    \author{A. Bonfanti\inst{1,2} \and S. Ortolani\inst{1,2} \and G. Piotto\inst{1,2} \and V. Nascimbeni\inst{1,2}}
    \institute{Dipartimento di Fisica e Astronomia, Università degli Studi di Padova, Vicolo dell'Osservatorio 3, I-35122 Padova, Italy
    \and
    Osservatorio Astronomico di Padova, INAF, Vicolo dell'Osservatorio 5, I-35122 Padova, Italy}
    \date{}

 
  \abstract
   {}
   {This article aims to measure the age of planet-hosting stars (SWP) through stellar tracks and isochrones computed with the \textsl{PA}dova \& T\textsl{R}ieste \textsl{S}tellar \textsl{E}volutionary \textsl{C}ode (PARSEC).}
   {We developed algorithms based on two different techniques for determining the ages of field stars: \emph{isochrone placement} and \emph{Bayesian estimation}. Their application to a synthetic sample of coeval stars shows the intrinsic limits of each method. For instance, the Bayesian computation of the modal age tends to select the extreme age values in the isochrones grid. Therefore, we used the isochrone placement technique to measure the ages of 317 SWP.}
   {We found that $\sim6\%$ of SWP have ages lower than 0.5 Gyr. The age distribution peaks in the interval [1.5, 2) Gyr, then it decreases. However, $\sim7\%$ of the stars are older than 11 Gyr. The Sun turns out to be a common star that hosts planets, when considering its evolutionary stage. Our SWP age distribution is less peaked and slightly shifted towards lower ages if compared with ages in the literature and based on the isochrone fit. In particular, there are no ages below 0.5 Gyr in the literature.}
   {}


   \maketitle
%

\section{Introduction}

Knowledge of the ages of stars with planets (SWP) is important for studying several aspects of the evolution of planetary systems, such as dynamical interactions among planets (see e.g. \cite{laughlin01}) and tidal effects induced by SWP (see e.g. \cite{paetzold04} \cite{barker09}). See \cite{hut80}, \cite{hut81} for a theoretical approach regarding tides. Moreover, SWP ages enable assessment of the evolution of the atmosphere of the hosted planets caused by chemical reactions occurring on the planets themselves and by the consequences of tidal stripping or other atmospheric loss processes. Knowledge of the stellar ages is also useful for selecting candidates for planet detections.

Most of the known SWP are main sequence G-K type stars that belong to the nearby disk field population. It is very well known that determining the ages of these stars is difficult because of the degeneracy of parameters and the slow evolution of the observational quantities. The current uncertainties are higher than the accuracy needed for these studies.

The age is not a direct observable, so its computation should use models or a combination of models and empirical relations. Methods based on isochrones from stellar evolutionary models are often used, but other methods are also applied, based on empirical relations, such as gyrochronology and activity indices. Asteroseismology will be a very promising technique when more specific data is available, and in specific cases, the chemical analysis (the so-called chemiochronology) can be applied as well. See \cite{soderblom10} for a broad review on this topic.

Most of the ages of SWP come from individual sources and different methodologies. A recent discussion of the ages of nearby field stars is presented in \cite{haywood13}, while specific analyses of SWP have been published by \cite{saffe05} and \cite{takeda07}. Finally \cite{brown14} presents a new study based on a new geometrical approach for interpolating the grids of isochrones, and he shows that the results from gyrochronology give systematically younger ages.

In this paper we focus on the ages derived by using isochrones. To determine the age of a stellar cluster using isochrones, it is necessary to put its stars on the Hertzsprung-Russel diagram (HRD) and evaluate --- among the isochrones having the metallicity of the cluster --- which isochrone best fits the layout of the stars on the diagram (\emph{isochrone fitting}). Instead, determining the age of a field star, in particular a MS star, is much more complex. The statistical treatment of the data plays a crucial role in the analysis, and it is necessary to face the problem of the degeneracy of parameters. Two different methods are typically applied: the \emph{isochrone placement} and the \emph{Bayesian estimation} (\cite{jorgensen05,dasilva06,pont04}). 

The isochrone placement technique consists in putting a star on the HRD together with its error bars in $\log{T_{\mathrm{eff}}}$ and $\log{L}$ and in properly selecting the best isochrone to account for the error box. Instead, the Bayesian estimation technique requires getting a posterior probability density function (\emph{pdf}) of the age of a star, assuming an a priori star formation rate, an a priori metallicity distribution, and an a priori initial mass function (IMF).

We note that sometimes statistical instruments such as Markov Chain Monte Carlo (MCMC) have been applied in the literature. MCMC tools are able to sample \emph{pdf}s (without obtaining an explicit expression for them), in order to infer parameters in Bayesian ambit. It can happen that useful parameters for computating stellar ages are recovered via MCMC tools and then used to compute the age through $\chi^2$-minimization-based methods. Algorithms where a preliminary Bayesian approach is followed by a frequentistic one are not strictly Bayesian.

The paper is organized in the following way. In §~\ref{sec:data} the input data and the isochrones we used are presented; in §~\ref{sec:methods} the implementation of isochrone placement and Bayesian estimation techniques is described; in §~\ref{sec:results} the results obtained are discussed; while §~\ref{sec:conclusion} reports a summary of our work.

\section{The data}\label{sec:data}
First of all, this section presents the input data that were used to test the reliability of the developed algorithms. The last two subsections are dedicated to describing the sample of SWP and of the theoretical models used to characterize the SWP. From here on, all the photometric parameters are expressed in the Johnson system.

\subsection{The 3.2 Gyr synthetic stars catalogue}
We built a catalogue of 927 synthetic stars located on an isochrone with initial [Fe/H]=0 and $\log{t}=9.5$ ($t\approx3.2$ Gyr). To each of them we attributed a distance $d$, its uncertainty $\Delta d$, and an uncertainty on $\log{g}$ $\Delta\log{g}$, generating random numbers from normal distributions with means $\bar{d}=135$ pc, $\overline{\Delta d}=0.1\bar{d}$, and $\overline{\Delta\log{g}}=0.1$ dex, respectively. The uncertainty on [Fe/H] is $\Delta\mathrm{[Fe/H]}=0.05$. All these values are typical of the stars belonging to the SWP catalogue that is described below.

\subsection{SWP Catalogue}
\begin{figure}
 \centering
 \includegraphics[width=\columnwidth]{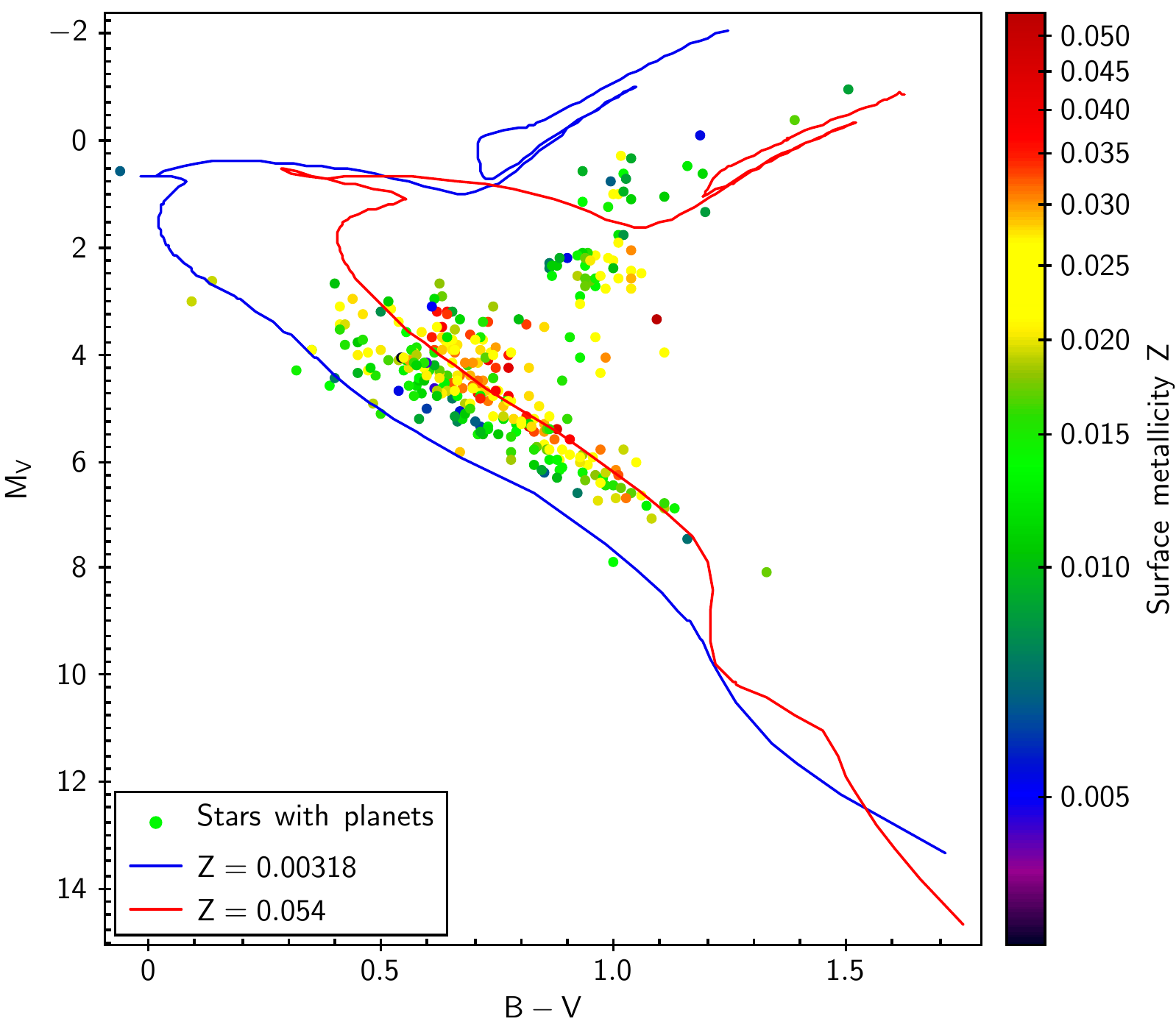}
 \caption{SWP on the CMD. The colour of the points is representative of the metallicity $Z$ of the stars. As reference, the 1-Gyr-isochrones corresponding to the extreme metallicity values of the sample are also represented.}
 \label{fig:CMD_SWP}
\end{figure}

All the planetary and stellar parameters of SWP were collected from \emph{The Site of California and Carnegie Program for Extra-solar Planet Search: Exoplanets Data Explorer}\footnote{\url{http://exoplanets.org/exotable/exoTable.html}} (\cite{wright11}). After having discarded binaries and stars without measurements of the apparent $V$ magnitude, $B-V$ colour index, parallax $\pi$ from Hipparcos, [Fe/H], and $\log{g}$, we considered the remaining 326 stars that constitute our SWP catalogue. These stars are represented on the CMD in Fig. \ref{fig:CMD_SWP} with the two 1-Gyr-isochrones corresponding to the minimum ($Z=0.00318$) and the maximum ($Z=0.054$) metallicity of the sample. Increasing the metallicity, the isochrones go towards redder colours. Some stars are located to the left of the $Z=0.00318$ 1-Gyr-isochrone. These are anomalous because they are expected on the right-hand side, considering their higher metallicity.

\subsection{The isochrones}\label{subsec:isochrone}
The theoretical models employed to determine the ages of the stars are the PARSEC\footnote{%
The web interface called CMD 2.5 input form is available at \url{http://stev.oapd.inaf.it/cgi-bin/cmd}} %
isochrones (version 1.0) by \cite{bressan12}, corresponding to the solar parameters listed in Table \ref{tab:SolarParam}.
\begin{table}
\centering
\caption{Solar parameters adopted by the isochrones.}
\label{tab:SolarParam}
\resizebox{\columnwidth}{!}{
\begin{tabular}{ccl}
\hline
Solar data		& 	Value				&	Source			\\
\hline
$Z_{\odot}$		&	0.01524				&	\cite{caffau08}		\\
$Z_{\mathrm{ini,}\odot}$&	0.01774				&	\cite{bressan12}	\\
$L_{\odot}$ [erg/s]	&	$(3.846\pm0.005)\cdot10^{33}$	&	\cite{guenther92}	\\
$R_{\odot}$ [km]	&	$695980\pm100$			&	\cite{guenther92}	\\
$T_{\mathrm{eff},\odot}$ [K]&	$5778\pm8$			&from $L_{\odot}$ and $R_{\odot}$\\
$M_{\mathrm{bol,}\odot}$ [mag]&4.770				&	\cite{girardi08}	\\
$BC_{\odot}$ [mag]	&	$-0.063$			&\multicolumn{1}{c}{interpolation}  		\\
$B-V_{\odot}$ [mag]	&	0.667				&\multicolumn{1}{c}{in the isochrones}	\\
$\log{g_{\odot}}$ [cgs]&	4.432				&\multicolumn{1}{c}{grid}		\\
$M_{V,\odot}$ [mag]	&	4.833				&from $M_{\mathrm{bol,}\odot}$ and $BC_{\odot}$ \\
$V_{\odot}$ [mag]	&	$-26.739$			&from $M_{V,\odot}$ and $d_{\oplus-\odot}$	\\
\hline
\end{tabular}
}
\end{table}
The different sequences of isochrones are identified by the metallicity of a star at the moment of its birth: $Z_{\mathrm{ini}}$. We used sequences spaced by 0.05 in $\log{t}$ (with $t$ in years) starting from $\log{t}=6$ up to $\log{t}=10.1$.

Considering the solar sequence of isochrones identified by $Z_{\mathrm{ini}}=Z_{\mathrm{ini,}\odot}=0.01774$ as reported by \cite{bressan12} (note that this value is different from the present one, which is $Z_{\odot}=0.01524$), the interpolation between the $\log{L/L_{\odot}}$ and $M_{\mathrm{bol}}$ tabulated values yields the correspondence $1L_{\odot}\rightarrow M_{\mathrm{bol}}=4.770$. We also computed the differences between the absolute bolometric and $V$ magnitudes given by the grid (which correspond to the bolometric corrections in the $V$ band $\mathrm{BC}=M_{\mathrm{bol}}-M_V$ adopted by the authors), and finally, we looked for the BC value that gave $T_{\mathrm{eff}}=T_{\mathrm{eff},\odot}$ and $L=L_{\odot}$. The interpolation gives a bolometric correction for the Sun $\mathrm{BC}_{\odot}=-0.063$. Finally, $\log{g_{\odot}}$ and $B-V_{\odot}$ come from the interpolation in the solar isochrone grid, as well.

As already said, the parameter that identifies a given sequence of isochrones is the metallicity $Z$, which is linked to [Fe/H] by an exponential relation\footnote{%
A reference relation between $Z$ and [Fe/H] is $\log{z}=\mathrm{[Fe/H]}+\log{(0.6369f_\alpha+0.3631)}-1.658$ proposed by \cite{straniero92}. It takes the possibility of $\alpha$-enhancement into account, where $\log{f_\alpha}=[\alpha/\mathrm{Fe}]$. (\ref{eq:metallicity}) is a reduced version of this equation considering $f_\alpha=1\Rightarrow[\alpha/\mathrm{Fe}]=0$, i.e. assuming a solar $\alpha$-enhancement. Moreover, the constant $-1.658$ is substituted with $-1.817$ so that $Z=Z_{\odot}=0.01524$ for [Fe/H]=0.}
, which assumes the form
\begin{equation}
Z=10^{\mathrm{[Fe/H]}-1.817}.
\label{eq:metallicity}
\end{equation}

\section{Age determination methods}\label{sec:methods}
\subsection{Preliminary considerations}\label{subsec:preliminary}

\begin{figure}
 \includegraphics[width=\columnwidth]{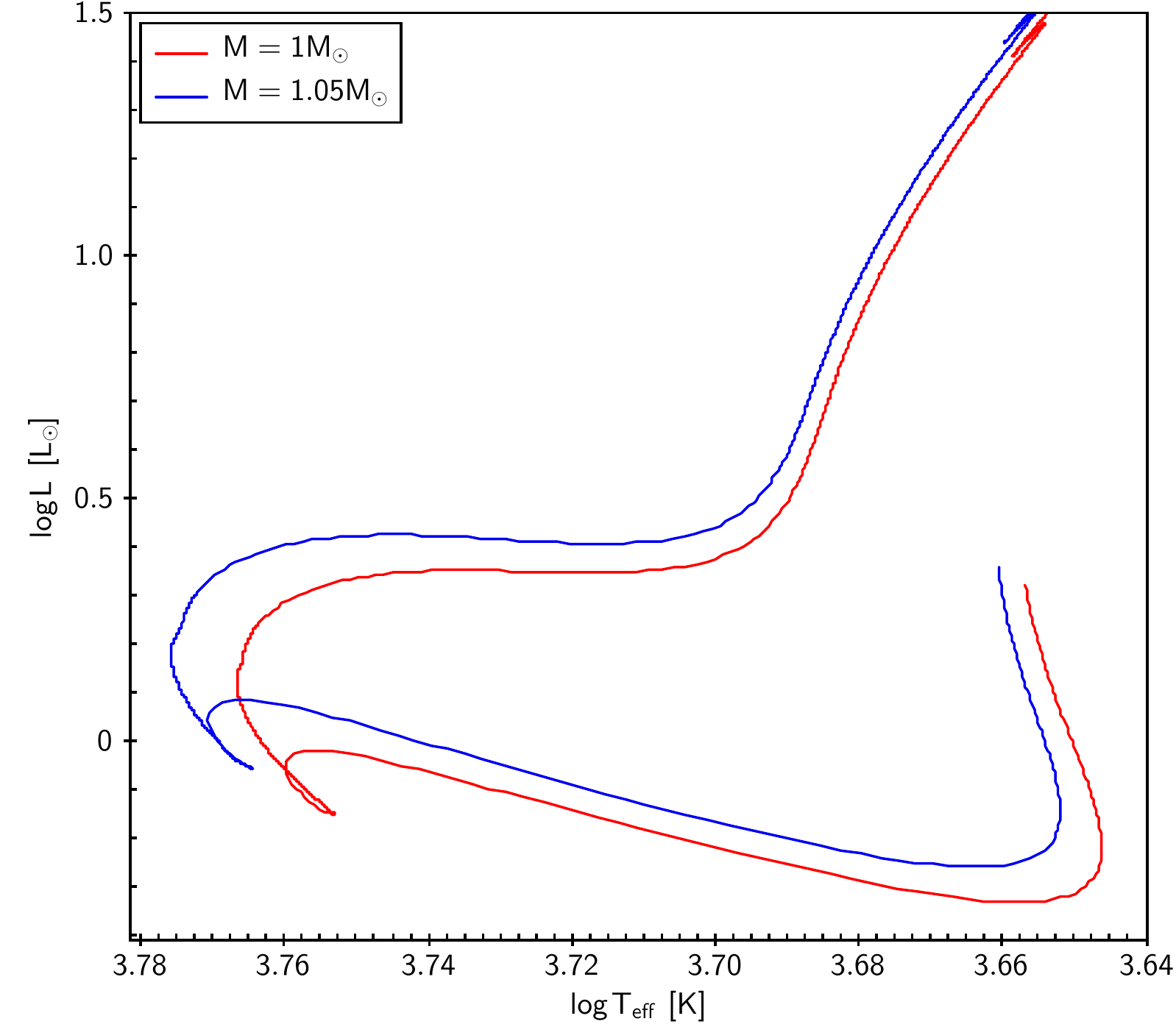}
 \caption{Solar metallicity evolutionary tracks of 1 and 1.05 $M_{\odot}$ star on the HRD. Their intersection point is representative of the degeneracy between stellar mass and age. See text for further details.}
 \label{fig:TracksM1-1.05}
\end{figure}

Deriving the ages of stars by making use of theoretical models requires knowledge of the stellar mass $M$. In fact, since isochrones corresponding to different age values can be very close on the HRD, $M$ can help in the selection of the one that fits the input data best. In some cases, there is even a degeneracy between the age $t$ and the mass $M$ of a given star. Figure \ref{fig:TracksM1-1.05}, which shows the solar metallicity evolutionary tracks of 1 and 1.05 $M_{\odot}$, clarifies the situation: the stellar parameters ($\log{T_{\mathrm{eff}}}$, $\log{L}$) = (3.7662, 0.0839) corresponding to the intersection of the two tracks are representative either of a 1 $M_{\odot}$ star with an age of 6.34 Gyr or of a 1.05 $M_{\odot}$ with an age of 26.8 Myr, so the knowledge of the stellar mass is fundamental to correctly establishing the evolutionary stage of any given star.

In the particular case where the comparison is between ages of a pre-main sequence (PMS) star and a MS star, it is also possible to remove the degeneracy considering that PMS stars are particularly active if compared with MS stars, and this implies that they have very high rotational velocities and cromospheric activity indices. Considering $\log{R'_{HK}}$ as reference index for the cromospheric activity, we took the age-activity relation by \cite{mamajek08} and we slightly shifted it, so that solar values adopted by the isochrones of Padova match the model. According to this relation $\log{R'_{HK}}=-4.48$ for an age $t=500$ Myr and $\log{R'_{HK}}=-4.27$ for $t=100$ Myr. The typical variation between the highest and lowest peaks in activity and the average level is $\sim0.2$ dex for a solar type star; in fact the present mean solar $\log{R'_{HK,\odot}}=-4.91$, while it was $\log{R'_{HK,\odot\mathrm{,Maunder}}}=-5.105$ during the Maunder minimum. To be conservative we assume that if
\begin{equation}
 \log{R'_{HK,\star}}<-4.65,
 \label{eq:Activity5}
\end{equation}
then the star has an age $t>500$ Myr, while if
\begin{equation}
 \log{R'_{HK,\star}}<-4.47,
 \label{eq:Activity1}
\end{equation}
then the star has an age $t>100$ Myr.

As mentioned before, another indicator of the activity of a star is its rotational velocity. According to the study of \cite{denissenkov10}, a star younger than 500 Myr at the very least has an angular velocity $\Omega_{\star}\gtrsim2.65\Omega_{e,\odot}$, while a star younger than 100 Myr has, at least, $\Omega_{\star}\gtrsim4\Omega_{e,\odot}$, where $\Omega_{e,\odot}=2.86\cdot10^{-6}$ rad/s is the present angular velocity of the solar envelope. In the absence of the $\log{R'_{HK,\star}}$ value, and assuming $\frac{4}{\pi}v\sin{i}$ as the mean probable rotational velocity of a star, we conclude that a star has an age $t>500$ Myr, if
\begin{equation}
 \Omega_{\star}<7.579\cdot10^{-6}\quad\mathrm{rad/s},
 \label{eq:Rotation5}
\end{equation}
while its age is greater than 100 Myr, if
\begin{equation}
 \Omega_{\star}<1.144\cdot10^{-5}\quad\mathrm{rad/s}.
 \label{eq:Rotation1}
\end{equation}

Besides many other parameters, the databases we examined often reported the stellar mass $M$, but all these parameters 
were derived by different authors following different calibration procedures. To make as few assumptions as possible and to produce input parameters that enter a picture that is self-consistent with the theoretical values reported by the isochrones, we decided to start from the values of
\begin{itemize}
\item visual magnitude $V$,
\item colour index $B-V$,
\item parallactic distance $d$,
\item metallicity [Fe/H], and
\item spectroscopic $\log{g}$,
\end{itemize}
which are available in the literature, and then to compute all the other needed parameters (i.e. the stellar mass $M$ using the stellar effective temperature $T_{\mathrm{eff}}$, luminosity $L$, and radius $R$), according to the calibrations that can be inferred from the values tabulated in the isochrones. 

The results are sensitive to the bolometric corrections (BCs). Several published tables of bolometric corrections are reported in the literature, but --- as pointed out by Torres (2010) --- values given by an author can differ noticeably from the ones given by another author depending on the arbitrary zero point (traditionally set using the Sun as reference) that each author adopts. Moreover, there is sometimes no internal consistency between $V_{\odot}$, $M_{\mathrm{bol,}\odot}$, and $\mathrm{BC}_{\odot}$.

For these reasons we obtained the correspondence between $M_V$ and $\log{L}$ by interpolating the values of the isochrone grids and inferring the BCs from the photometric values tabulated in the isochrones. Assuming an internal uncertainty on the apparent bolometric magnitude equal to 0.03 mag, the uncertainty $\Delta L$ is associated to $L$ through error propagation. The correspondence between $B-V$ and $\log{T_{\mathrm{eff}}}$ is obtained from the isochrones as well. An internal uncertainty of 1\% is attributed to the resulting $T_{\mathrm{eff}}$ value.

From the parameters just derived, we can now compute the estimates of $R$ and $M$ that are used as input data in the isochrone placement technique:
\begin{equation}
R=\sqrt{\frac{L}{\left(\frac{T_{\mathrm{eff}}}{5778}\right)^4}}
\label{eq:radius}
\end{equation}
\begin{equation}
M=\frac{g}{10^{4.43}}R^2.
\label{eq:mass}
\end{equation}
In these equations $L$ and $M$ are in solar unities, $T_{\mathrm{eff}}$ is in K, and $g$ is in cm/s$^2$. After applying the error propagation, all the data are accompanied by the respective uncertainties.

Another aspect that has been investigated is the temporal evolution of the surface stellar metallicity $Z$ due to atomic diffusion. The interaction between different chemical species leads to a surface depletion of elements heavier than hydrogen, which sink downwards. The characteristic timescale for the diffusion of an element is (\cite{chaboyer01})
\begin{equation}
 \tau_{\mathrm{diff}}\simeq K\frac{M_{\mathrm{CZ}}}{MT_{\mathrm{CZ}}^{3/2}}
 \label{eq:diffusion}
\end{equation}
where $M$ is the stellar mass, $M_{\mathrm{CZ}}$ the mass of the surface convective zone, $T_{\mathrm{CZ}}$ the temperature at the base of the convective zone, and $K$ a constant referring to the chemical element being taken into account. The direction of this process is parallel to the temperature and pressure gradients, while it is antiparallel to the chemical concentration gradient. For further information see \cite{burgers69} and \cite{chapman70}.

\begin{figure}
 \includegraphics[width=\columnwidth]{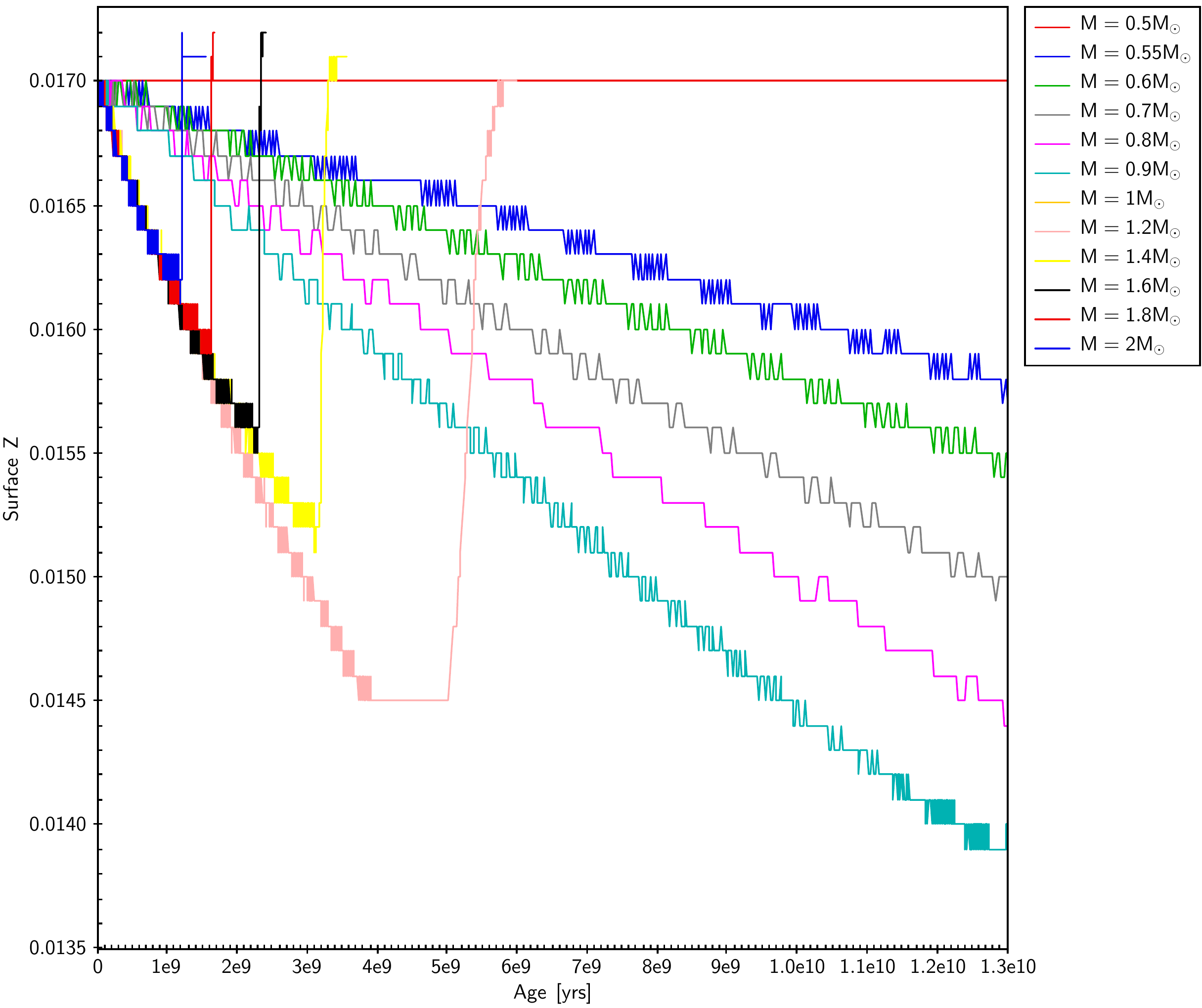}
 \caption{Evolution with time of the surface metallicity $Z$ for stars of different masses characterized by the same initial metallicity $Z_{\mathrm{ini}}=0.017$. The ruggedness of the curves is due to the discrete steps in the model, and it has been smoothed while implementing our routines.}
 \label{fig:Zevo}
\end{figure}

We considered several stellar evolutionary tracks, kindly provided by Leo Girardi, which illustrate the evolution of stars of different masses and initial metallicities on the HRD. Such tracks give the surface Z in correspondence of any given age, so we built up different $Z_{k,l}=Z_{k,l}(t)$ functions depending on the stellar initial metallicity and mass (identified by the subscripts $k$ and $l$, respectively). We observed that atomic diffusion is not negligible for stars having masses between 0.5 and 2 $M_{\odot}$. In this mass range, the surface metallicity decreases with the time during the MS phase and then increases as the envelope convection deepens, becoming constant once the initial metallicity value $Z_{k,l}(0)$ is reached. As a consequence, the initial decrease followed by the increase in $Z_{k,l}(t)$ spans a shorter time scale for higher masses, where the evolution is faster. An example of the temporal evolution of $Z$ for stars of different masses, but characterized by the same initial 
metallicity, is shown in Fig \ref{fig:Zevo}.

The evolution of the atmospheric chemical composition has a non-negligible effect. For instance, if we take a solar star (i.e. with the current luminosity and temperature of the Sun) with a present metallicity equal to $Z_{\odot}=0.01524$ and employ the sequence of isochrones characterized by such metallicity without taking into account that the metallicity value that identifies the isochrones is the one that a star had when it was born and not the present one, then we get an age $t=5.1\pm2.8$ Gyr instead of the $t=4.5\pm0.1$ Gyr that we obtain if we adopt the initial metallicity value $Z_{\mathrm{ini,}\odot}=0.01774$ to select the isochrones.

\begin{figure}
 \includegraphics[width=\columnwidth]{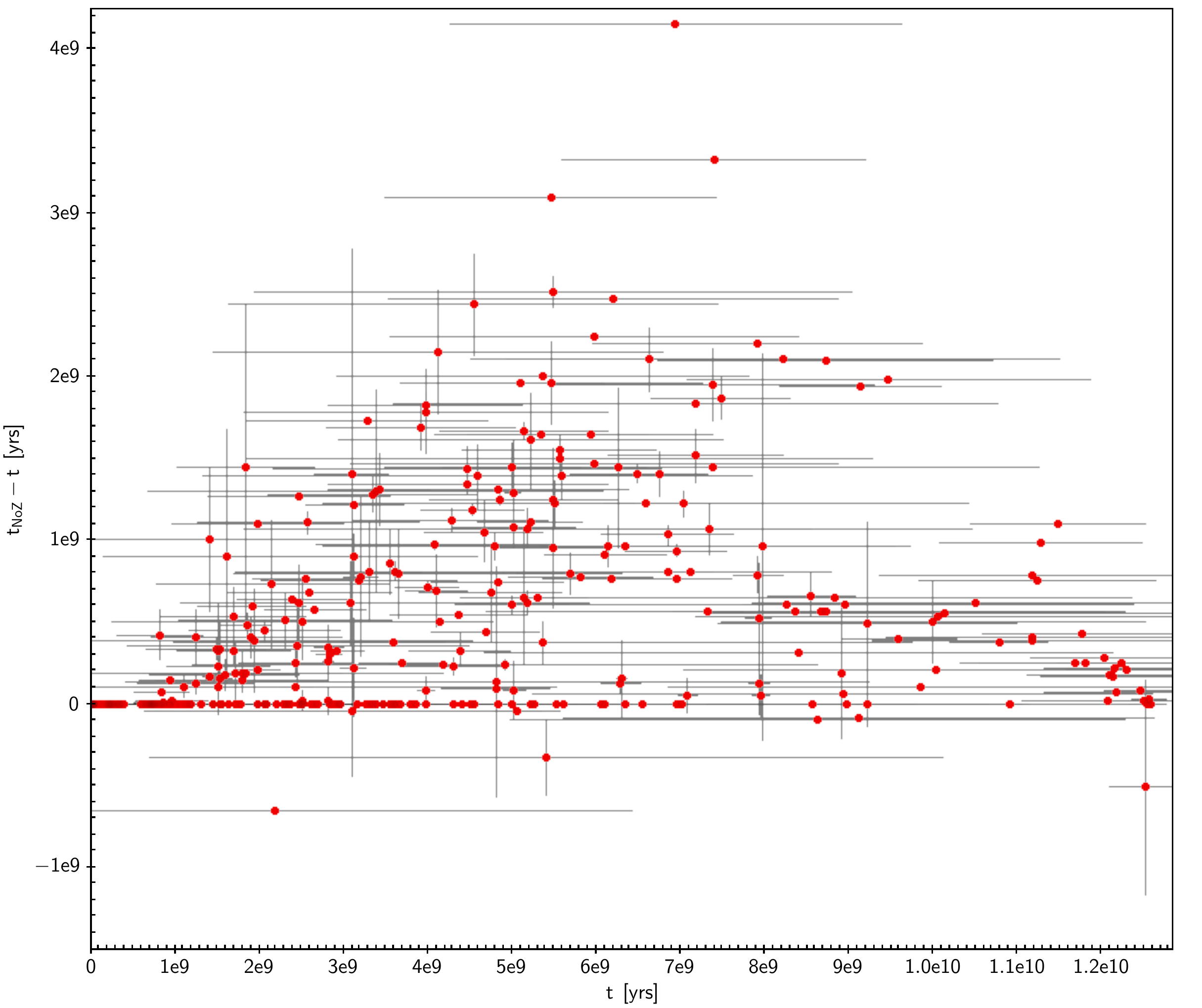}
 \caption{SWP ages computed through the isochrone placement technique. The difference $t_{\mathrm{noZ}}-t$ between the ages $t_{\mathrm{noZ}}$ computed without taking the surface metallicity $Z$ evolution into account and the ages $t$ computed considering such an effect, is plotted versus $t$. If the algorithm does not take into account that the sequences of isochrones are identified by the initial stellar metallicity, while we only know the present value, then the ages obtained are biased towards older values, especially for intermediate-age stars.}
 \label{fig:CfrSWPageZevo}
\end{figure}

Adopting the present-day atmospheric chemical composition in selecting the isochrones for the age computation of field stars generally produces a result that is slightly biased towards older ages, especially for intermediate-age stars, as shown in Fig. \ref{fig:CfrSWPageZevo}. In fact, higher metallicity isochrones are redder than lower ones for every age, and in the MS the older a star, the redder it is. Since considering the initial chemical composition of a star implies selection of an equally or higher metallic grid of isochrones, we expect $t_{\mathrm{noZ}}-t\ge0$ for MS stars. Among pre-MS isochrones, instead, older isochrones are bluer for a given metallicity: in this case, we expect negative differences. The three stars in Fig. \ref{fig:CfrSWPageZevo} with evident $t_{\mathrm{noZ}}-t<0$ are peculiar because their colours are too red for their luminosity and metallicity. They are located in a region where there are both old and pre-MS isochrones. Furthermore, they do not have activity 
indices to disentangle between young and old ages; in these conditions, the code may give $t_{\mathrm{noZ}}<t$. This result could be eliminated by iterating the isochrone placement twice and using in the second iteration only those isochrones that differ $\pm\Delta t$ from the age value $t$ recovered by the first iteration. We are considering implementating this in the near future. 

\subsection{Isochrone placement}
The input data characterizing each star for which we want to establish the age are listed below:
\begin{center}
\resizebox{\columnwidth}{!} {
\begin{tabular}{*{14}{c}}
$Z$	&	$V$	&	$\mathrm{BC}$	&	$B-V$	&	$d$	&	$\Delta d$	&	$T_{\mathrm{eff}}$	& $\Delta T_{\mathrm{eff}}$	&	$L$	&	$\Delta L$	&	$g$	&	$\Delta g$	&	$M$	&	$\Delta M$.	\\
\end{tabular}
}
\end{center}
To determine the age and the other parameters, such as effective temperature, luminosity, gravity, and mass, of a given star according to the theoretical models, we first considered the sequence of isochrones \emph{Is} characterized by  the present metallicity of the star $Z_{\star}$. If $\log{R'_{HK}}$ or $v\sin{i}$ of a star was available (which happens for the 94\% of the stars that belong to the SWP catalogue) and relations (\ref{eq:Activity5}) or (\ref{eq:Rotation5}) were satisfied, we discarded all the rows reporting ages lower than 500 Myr from the sequence \emph{Is}. If, instead, relations (\ref{eq:Activity5}) and (\ref{eq:Rotation5}) did not hold, but relations (\ref{eq:Activity1}) or (\ref{eq:Rotation1}) were satisfied, we discarded all the age values available in the isochrone grids lower than 100 Myr (activity cleaning). 

Then, for each isochrone in the \emph{Is} sequence, we considered the point on the CMD with the minimum distance from the given star and computed all its corresponding theoretical values by interpolation through the isochrone grid values. We thus built our reduced grid of isochrones \emph{Is'}. After that, we developed the following procedure:
\begin{enumerate}
\item For each row $i = 1, 2, \dots, \bar{n}$ of \emph{Is'} (from here on, the subscript $i$ will always represent the row index of \emph{Is'}), multiply the theoretical value reported in the $i^{\mathrm{th}}$ row of \emph{Is'} (we refer to it as the generic variable $X_{i}$ that indicates, time by time, $T_{\mathrm{eff}}$, $L$, $M$, or $g$) by the bidimensional Gaussian distribution (window function):
  \begin{equation}
  \mathcal{G}(L_i, T_{\mathrm{eff},i})= \frac{e^{-\frac{1}{2}\left[\left(\frac{\log{T_{\mathrm{eff},i}}-\log{T_{\mathrm{eff}}}}{\Delta\log{T_{\mathrm{eff}}}}\right)^2+\left(\frac{\log{L_i}-\log{L}}{\Delta\log{L}}\right)^2\right]}}{2\pi \Delta\log{T_{\mathrm{eff}}} \Delta\log{L}}.
  \label{eq:Gauss}
  \end{equation}
	
In this way we can consider that the probability that a given stellar parameter corresponds to the value reported by a certain row of \emph{Is'} decreases with increasing distance on the HRD between the star and the isochrone itself, but --- at the same time --- the isochrones falling out of the error bars in $\log{T_{\mathrm{eff}}}$ and $\log{L}$ are not discarded definitely.
\item Compute the weight
\begin{equation}
\scriptstyle
p_{i}=\left[\left(\frac{\log{L}-\log{L_i}}{\Delta\log{L}}\right)^2+\left(\frac{\log{T_\mathrm{eff}}-\log{T_{\mathrm{eff,}i}}}{\Delta\log{T_{\mathrm{eff}}}}\right)^2+\left(\frac{M-M_i}{\Delta M}\right)^2+ \left(\frac{\log{g}-\log{g_i}}{\Delta\log{g}}\right)^2\right]^{-1}
\end{equation}

that must be attributed to $X_i$, so that the similarity between the stellar and the theoretical $M$ and $\log{g}$ values is also taken into account. The greater the likeness between the stellar and the theoretical data, the bigger $p_i$.
\end{enumerate}
Clearly, in the previous equations all the data with the subscript $i$ are taken from the isochrones, while the others are the input stellar parameters.

Through weighted means, it is now possible to compute the age of a given star and its temperature, luminosity, mass, and gravity according to the Padova evolutionary models. According to what has just been described, the generic stellar parameter expressed by $X_{\star}$ results in:

\begin{equation}
X_{\star}=\frac{\displaystyle\sum_{i=1}^{\bar{n}}{X_{i}\mathcal{G}(L_i, T_{\mathrm{eff},i})p_i}}{\displaystyle\sum_{i=1}^{\bar{n}}{\mathcal{G}(L_i, T_{\mathrm{eff},i})p_i}}.
\label{eq:X.is}
\end{equation}

The corresponding uncertainty is given by

\begin{equation}
\Delta X_{\star}=\sqrt{\frac{\displaystyle\sum_{i=1}^{\bar{n}}{(X_{i}-X_{\star})^2\mathcal{G}(L_i, T_{\mathrm{eff},i})p_i}}{\displaystyle\sum_{i=1}^{\bar{n}}{\mathcal{G}(L_i, T_{\mathrm{eff},i})p_i}}}.
\label{eq:DeltaX.is}
\end{equation}

Making use of these first guesses for $t_{\star}$ and $M_{\star}$, among the $Z_{k,M_{\star}}$ functions describing the evolution in metallicity of a star with $M=M_{\star}$, we looked for the one where $Z(t_{\star})=Z_{\star}$, from which we recovered the initial metallicity $Z_{\mathrm{ini,}\star}$ that a star of age $t_{\star}$ had at its birth if the present metallicity is $Z_{\star}$. After that, we considered the sequence of isochrones corresponding to the just estimated metallicity $Z_{\mathrm{ini,}\star}$ and we iterated all these operations until a convergence in the stellar age value was reached.

\subsection{Bayesian estimation}
As presented by \cite{jorgensen05}, determining stellar ages from isochrones requires a comparison between observational and theoretical data, according to the stellar evolutionary model adopted. If the model-relevant parameters are collected in a vector \textbf{p}, while the observational data are collected in a vector \textbf{q}, the theoretical model gives a map from the parameter space \textbf{p} to the data space \textbf{q}. Determining stellar ages represents the inverse problem, i.e. finding a map from \textbf{q} to \textbf{p}; here we considered $\mathbf{p}=(\tau, Z, m)$ (where $\tau$ is the age and $m$ is the mass) and $\mathbf{q}=(\mathrm{[Fe/H]}, \log{T_{\mathrm{eff}}}, \log{L})$.

In Bayesian statistics, the parameters that have to be estimated (in our case $\tau$, $Z$ and $m$) are treated as random variables, and their posterior (joint) probability density function is
\begin{equation}
f(\tau, Z, m)\propto f_0(\tau, Z, m)\mathcal{L}(\tau, Z, m)
\label{eq:fBayes}
\end{equation}
where $f_0$ is the prior probability density function and $\mathcal{L}$ the likelihood function. The value given by $f(\tau, Z, m)$d$\tau$d$Z$d$m$ represents the fraction of stars with age inside [$\tau$, $\tau$ + d$\tau$], metallicity inside [$Z$, $Z$ + d$Z$] and mass inside [$m$, $m$ + d$m$]. The constant of proportionality must be chosen so that $\int\int\int f(\tau, Z, m)$d$\tau$d$Z$d$m$ = 1.
The integration of $f$ with respect to $Z$ and $m$ gives $f(\tau)$, which is the posterior pdf that a star has the age $\tau$. Assuming the mode as the statistical index that synthesizes the function, the best estimate for the age is the value that maximizes $f(\tau)$. Other plausible choices are those referring to the most probable age of a star considering the mean of the pdf (corresponding to the centroid of the area under $f(\tau)$) or the median, which, instead, is the value that bisects the area under $f(\tau)$. As already pointed out by \cite{jorgensen05} --- who considered a sample of 2968 synthetic stars in order to evaluate the best criterion to assess age in Bayesian statistics --- the mean and the median suffer the bias of attributing an age that is in the centre of the sequence of age values reported by the isochrone grid employed. On the other hand, the mode tends to assign the extreme age values of the isochrone grid to the stars: in particular, selection of extreme ages 
arises for 2031 stars over the 2968 of their entire sample, which corresponds to a frequency of $\sim70\%$. In § \ref{sec:results} we confirm the behaviour of these three statistical indices by applying the Bayesian statistics to the stars of both the 3.2 Gyr catalogue and the SWP catalogue.

Assuming independent Gaussian observational errors $\sigma_n^{\mathrm{obs}}$ for each $q_n^{\mathrm{obs}}$, the likelihood function is given by
\begin{equation}
\mathcal{L}(\tau, Z, m)=\prod_{n=1}^3\frac{1}{\sqrt{2\pi}\sigma_n^{\mathrm{obs}}}\cdot\exp(-\chi^2/2)
\label{eq:likelihood}
\end{equation}
where
\begin{equation}
\chi^2=\sum_{n=1}^3\left(\frac{q_n^{\mathrm{obs}}-q_n(\tau,Z,m)}{\sigma_n^{\mathrm{obs}}}\right)^2.
\label{eq:chiquadro}
\end{equation}

Following the suggestion of \cite{jorgensen05}, we assumed
\begin{equation}
f_0(\tau, Z, m)=\psi(\tau)\phi(Z)\xi(m)
\label{eq:f0Bayes}
\end{equation}
where $\psi(\tau)$ is the a priori star formation rate (SFR), $\phi(Z)$ the a priori metallicity distribution, and $\xi(m)$ the a priori IMF. As one of our purposes is that of studying possible evolutionary peculiarities of our stellar samples, we assume that the prior SFR $\psi(\tau)$ is flat. Regarding the metallicity prior $\phi(Z)$, the spectroscopically determined [Fe/H] values we employed are very reliable (for example, $\sim85\%$ of the stars belonging to the SWP catalogue has $\sigma_{\mathrm{[Fe/H]}}<0.08$ dex, with the majority of them that having $\sigma_{\mathrm{[Fe/H]}}=0.03$ dex) and, since for very high accuracy metallicities (say $\sigma_{\mathrm{[Fe/H]}}\sim0.05$ dex) the likelihood function acts as a narrow window function that substantially suppresses the contribution of $\phi(Z)$, we consider a flat $\phi({Z})$ (see \cite{pont04} for a detailed discussion about the choice of the prior distributions). Assuming also that $\xi(m)=m^{-2.7}
$, which is representative of the empirical IMF at around 1 $M_{\odot}$ (\cite{kroupa93}), it is possible to obtain $f(\tau)$ as
\begin{equation}
f(\tau)\propto G(\tau)=\int\int \mathcal{L}(\tau,Z,m)\,\xi(m)\,\mathrm{d}m\,\mathrm{d}Z.
\label{eq:ftauBayes}
\end{equation}

Implementing the algorithm of the Bayesian determination of age, we first check that the condition
\begin{equation}
\min \chi_i^2<\chi_{0.99}^2
\label{eq:chi2condition}
\end{equation}
is satisfied; $\chi_{0.99}^2$ is the $99^{\mathrm{th}}$ percentile value of the chi-square distribution that in our case is 11.345, because we have three degrees of freedom (d.o.f). Since the probability that $\chi^2\le11.345$ for a 3 d.o.f. chi-square distribution is 99\%, relation (\ref{eq:chi2condition}) states that we do not evaluate ages for stars whose input parameters have less than 1\% probability (according to chi-square distribution) to actually be those measured. These stars are characterized by data points that are far away from any isochrone, so their pdfs turn out to be meaningless.

If the preliminary condition given by (\ref{eq:chi2condition}) is satisfied, then we numerically evaluate $G(\tau)$ considering a set of sequences of isochrones taken at constant steps of 0.05 dex in [Fe/H] within the interval whose bounds are empirically fixed at $\pm3.5\sigma_{\mathrm{[Fe/H]}}$ from the stellar metallicity, where $\sigma_{\mathrm{[Fe/H]}}$ is the uncertainty on [Fe/H]. Let $m_{jkl}$ be the initial mass value read at line $l$ of the isochrones grid of age $\tau_j$ and metallicity $Z_k$, then
\begin{equation}
G(\tau_j)=\sum_k\sum_l \mathcal{L}(\tau_j,Z_k,m_{jkl})\xi(m_{jkl})(m_{jkl+1}-m_{jkl}).
\label{eq:ftaujBayes}
\end{equation}

Once we have obtained the vector $G(\tau)$ of components $G(\tau_j)$, we find the component that assumes the maximum value (say $G(\tau_{\bar{j}})$) and divide each component by $G(\tau_{\bar{j}})$, obtaining the normalized function $\tilde{G}(\tau)$. After we have smoothed $\tilde{G}$ through a polynomial interpolation, the most probable age $\hat{\tau}$ attributed to the star is the value that maximizes $\tilde{G}$ (modal value). We also compute the mean age as the age coordinate of the centroid of the area under $\tilde{G}$ and the median age as the age value that bisects the area under $\tilde{G}$. 

As described by \cite{jorgensen05}, it is possible to prove that $\tilde{G}_{\mathrm{lim}}=0.61$ sets the 68\% confidence level of $\hat{\tau}$, so we provide a 68\% confidence interval $[\tau_1, \tau_2]$ to be the shortest interval such that $\tilde{G}(\tau)<0.61$ outside it.\footnote{%
Since $f(\tau)$ can have multiple maxima, it may happen that $f(\tau)$ is locally below 0.61 inside $[\tau_1, \tau_2]$} %

\section{Discussion of the results}\label{sec:results}
\subsection{Isochrone placement vs. Bayesian estimation}

The first aim of the paper is to compare the reliability of the developed algorithms in the computation of ages of field stars. To reach this goal, we applied these algorithms taking the 3.2 Gyr synthetic stars one by one. Then we checked the correspondence ($B-V$,$M_V$) $\rightarrow$ ($\log{T_{\mathrm{eff}}}$, $\log{L}$) representing each star on both the colour magnitude diagram (CMD) and the HRD. The calibration is correct if a star has the same relative position with respect to the same reference isochrones on both the two diagrams. Finally, we checked to what extent the global age distribution of the entire sample is consistent with the expected age of 3.2 Gyr attributed to the stars a priori. The calibration between observational and theoretical parameters is perfect except for the RGB region, where the intersection between different isochrones makes it difficult; this will have consequences on the age determination.

\begin{figure}
 \includegraphics[width=\columnwidth]{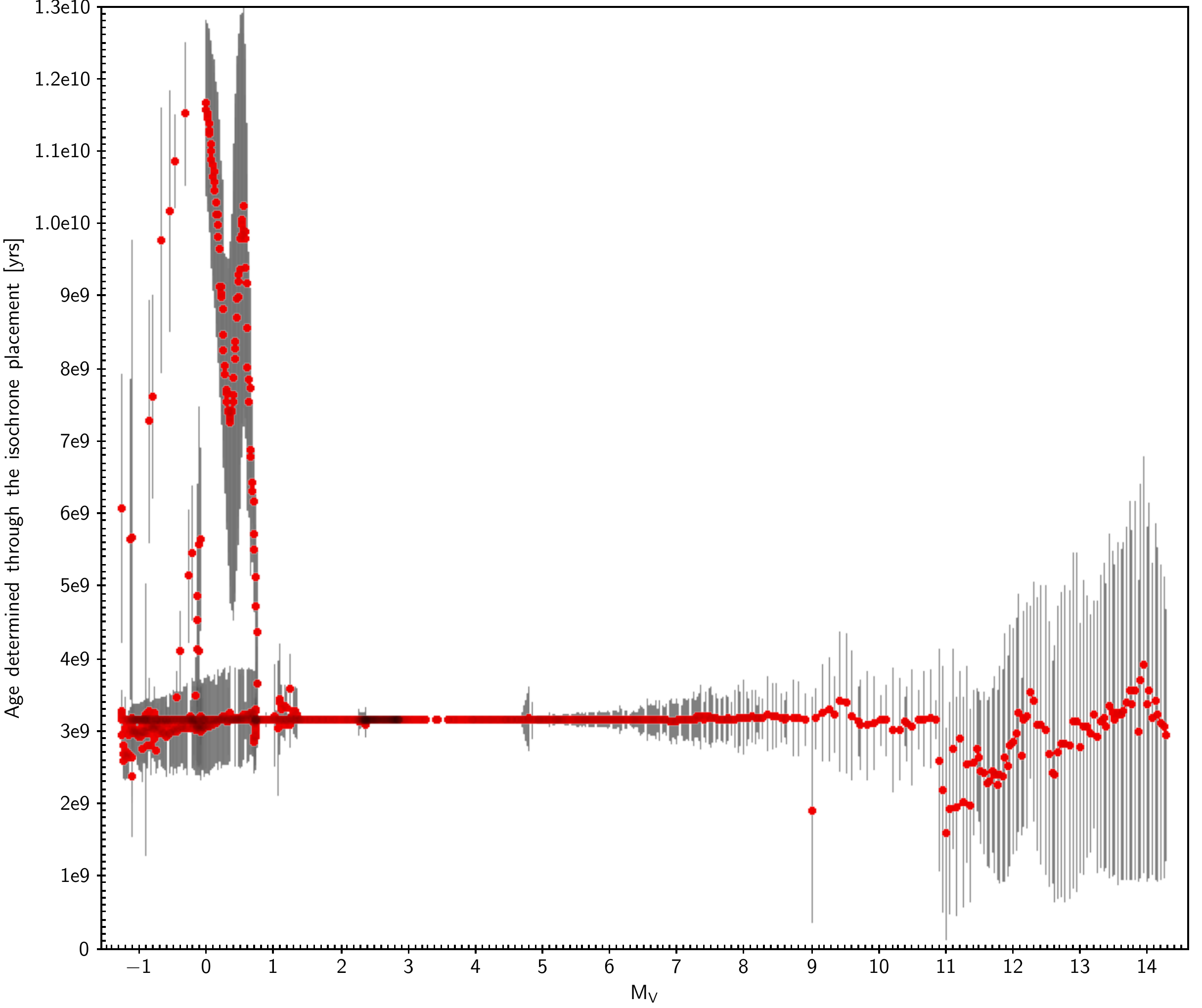}
 \caption{Ages of the 3.2 Gyr synthetic stars determined through the isochrone placement technique plotted versus their absolute magnitude $M_V$.}
 \label{fig:AgeVsV}
\end{figure}

In fact, the representation of the age of these stars determined with the isochrone placement technique versus their absolute magnitude $M_V$ (Fig. \ref{fig:AgeVsV}) shows that --- as expected --- the majority of the stars fall in the horizontal region between 3 and 3.5 Gyr, but some deviations occur for very low and very high magnitudes. The ages of some very bright stars (in the RGB phase) can be imprecise because of the difficulties linked to the calibration between observational and theoretical stellar parameters owing to the particular shape of the isochrones in that region. Instead, the errors for the ages obtained for the faintest stars deals with the intrinsic difficulties in estimating the ages of low MS stars, even if the calibration from the CMD to the HRD is well done. Just to summarize, the algorithm implementing the isochrone placement technique gives reliable ages, except for
\begin{itemize}
 \item some stars in the RGB phase, whose ages could also be completely wrong;
 \item low MS stars, with the obtained ages that can differ up to 50\% from the correct value.
\end{itemize}

We decided to further analyse the ages of the synthetic stars belonging only to the MS (i.e. stars having $4<M_V<8$), because they are the most common one amongst the SWP. In addition, since extremely young ages can be discarded considering the stellar activity, the ages are obtained by removing the theoretical isochrones with ages lower than 500 Myr from the fit procedure.

\begin{figure}
\includegraphics[width=\columnwidth]{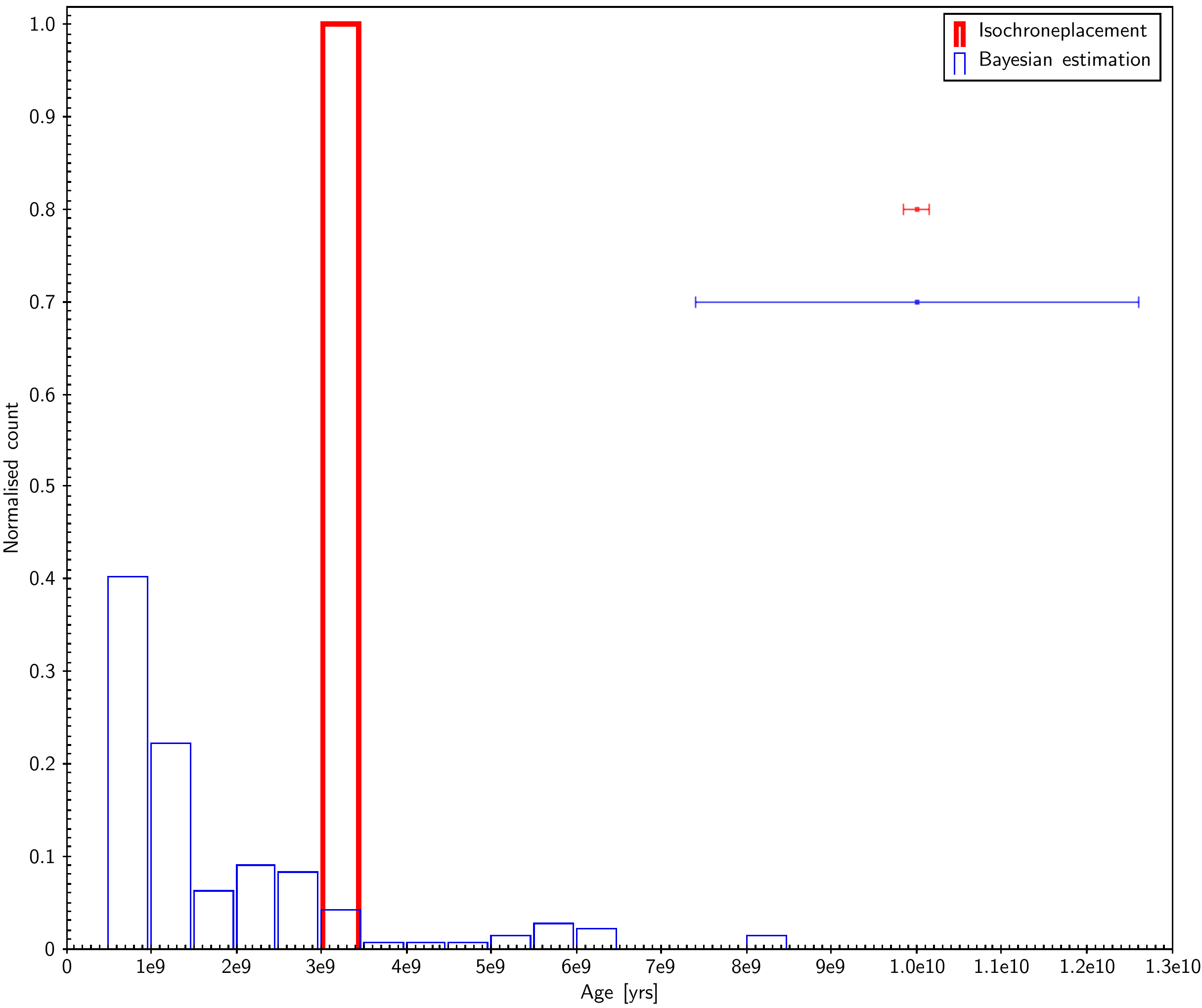}
\caption{Isochrone placement vs. modal Bayesian ages. Age distribution of the 3.2 Gyr synthetic stars having $4<M_V<8$.}
\label{fig:DistrEta3}
\end{figure}

\begin{figure}
 \includegraphics[width=\columnwidth]{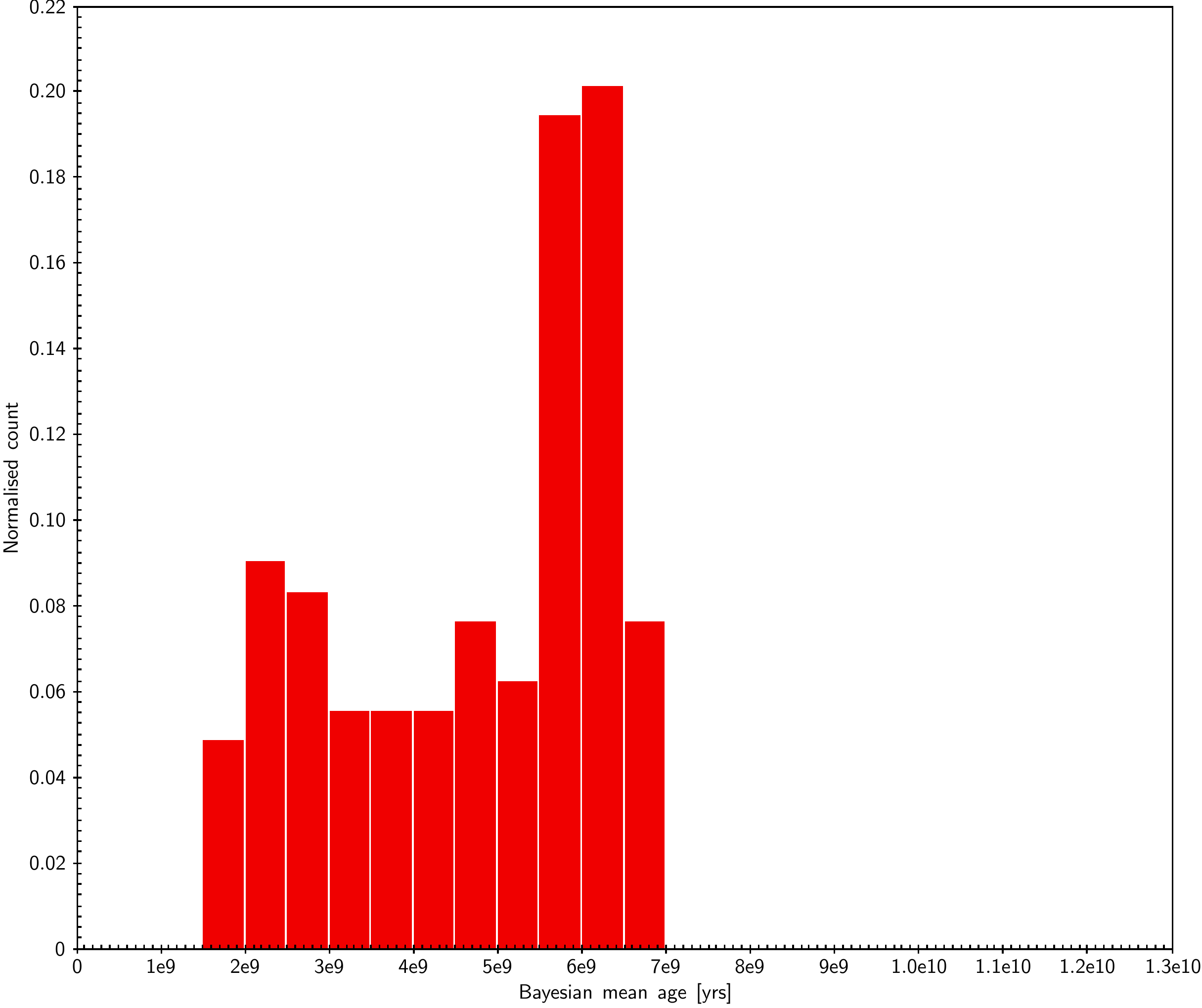}
 \caption{Bayesian estimation. Mean age distribution of the 3.2 Gyr synthetic stars.}
 \label{fig:DistrEta3Mean}
\end{figure}

\begin{figure}
 \includegraphics[width=\columnwidth]{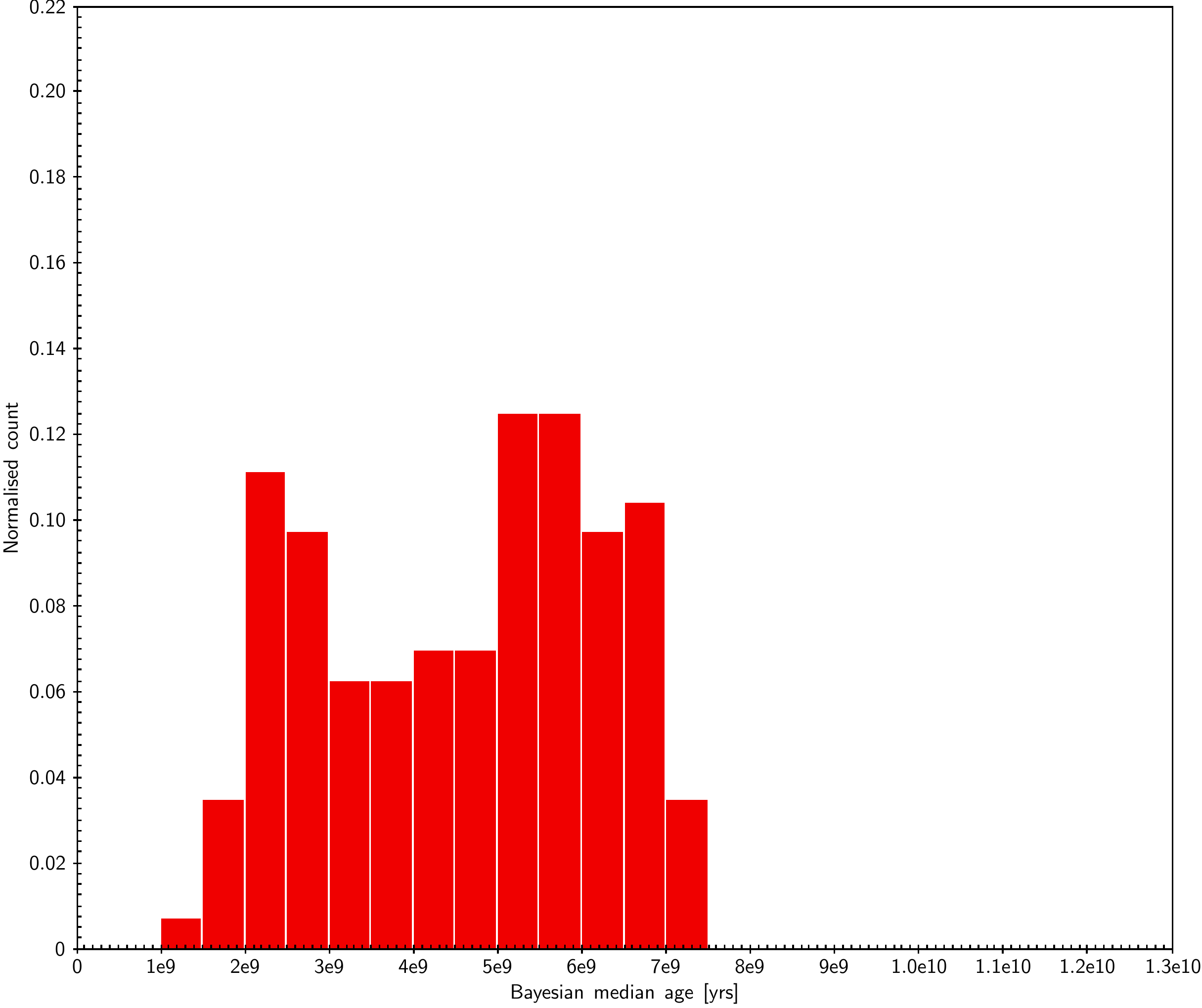}
 \caption{Bayesian estimation. Median age distribution of the 3.2 Gyr synthetic stars.}
 \label{fig:DistrEta3Median}
\end{figure}

The histogram in Fig. \ref{fig:DistrEta3} shows a comparison between the age distributions computed with the isochrone placement technique and the Bayesian estimation, which adopts the mode as synthesis index. It also reports the mean error bars associated to the results. Figures \ref{fig:DistrEta3Mean} and \ref{fig:DistrEta3Median} show the age distributions of the same synthetic stars as were obtained through the Bayesian estimation technique that adopts the mean and the median as synthesis indices, respectively.


The isochrone placement turns out to be the most reliable technique, since all the stars fall in the bin [3, 3.5] Gyr (which contains the value of age of 3.2 Gyr a priori attributed to these stars) with a typical uncertainty in the age determination of $\sim0.15$ Gyr, corresponding to an error of $\sim5\%$.

On the other hand, using the Bayesian statistics, the distribution derived from the modal age values presents the main peak in correspondence of [0.5, 1] Gyr. This is the outmost interval of the isochrone grid, and it is representative of an age that is lower than the correct one. Finally, the distribution obtained using the mean and the median as synthesis indices of the Bayesian pdfs are quite similar, essentially spanning an age range from 1 to 7 Gyr. The determination of ages through the mean or the median is therefore not very accurate, and confirming the conclusion by \cite{jorgensen05} and by \cite{takeda07}, these statistical indices tend to select ages in the middle of the sequence of age values reported by the isochrone grids, centralizing the distribution. 

\begin{figure}
 \centering
 \includegraphics[width=\columnwidth]{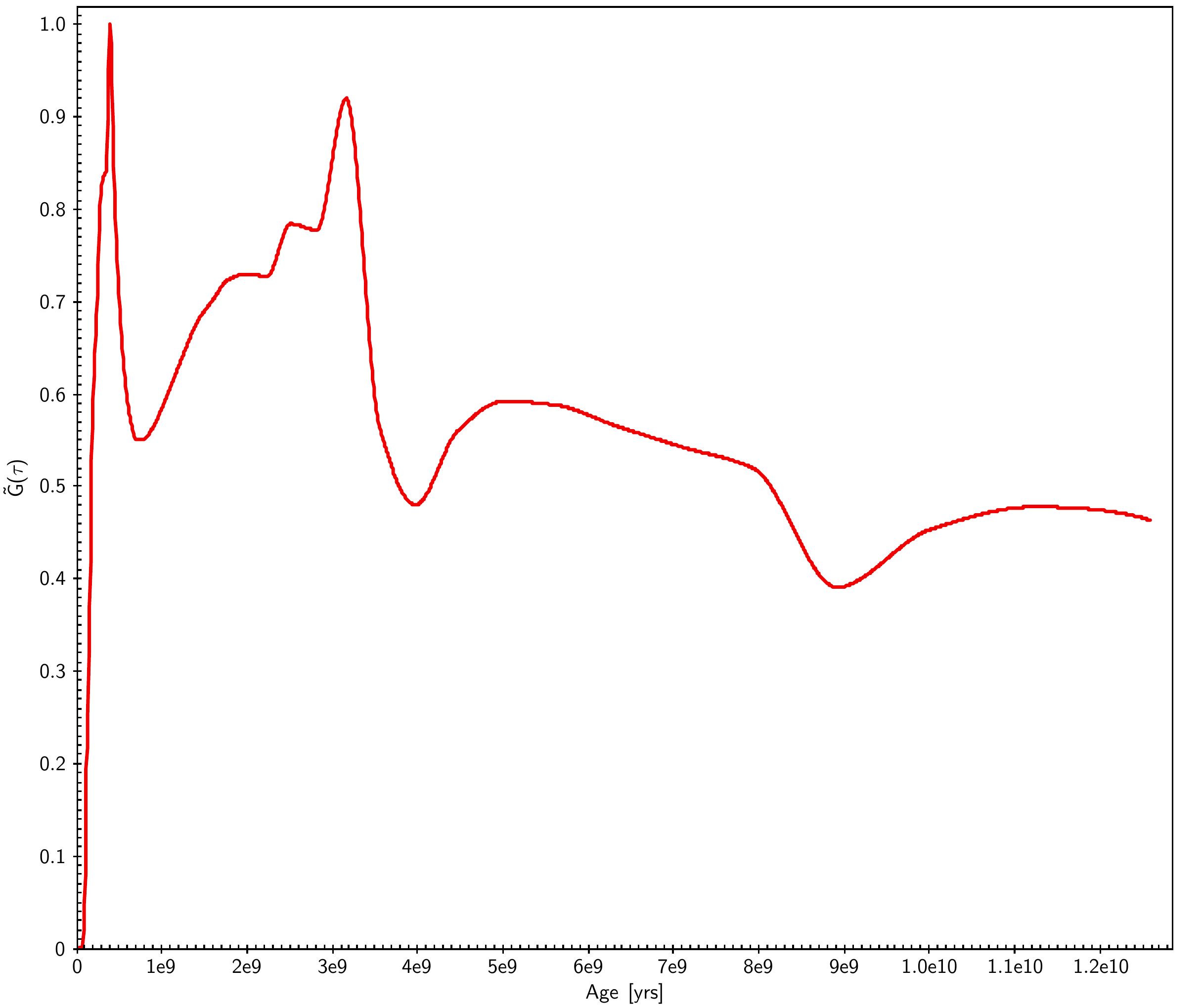}
 \caption{Bayesian $\tilde{G}(\tau)$ of a 3.2 Gyr synthetic star, which is representative of the presence of spurious peaks corresponding to low age values.}
 \label{fig:Bayespdf}
\end{figure}

The mode is definitely the indicator to be preferred and its tendency to select the extremes in age values in the isochrone grids could be partially mitigated operating a proper numerical filtering. In fact, in some cases --- an example of one of them is shown in Fig. \ref{fig:Bayespdf} --- the Bayesian pdf shows peaks corresponding to very low ages. Dealing with synthetic stars, in this context we can recognize such peaks as spurious, and we realize that they hide the presence of the peak centred at $\sim3$ Gyr, which indicates the correct age. However, if no a priori indication is given about the ages of stars (the ordinary situation if the scientific aim is to determine of the ages of field stars), it is not possible to select the true peak.

\begin{figure}
 \centering
 \includegraphics[width=\columnwidth]{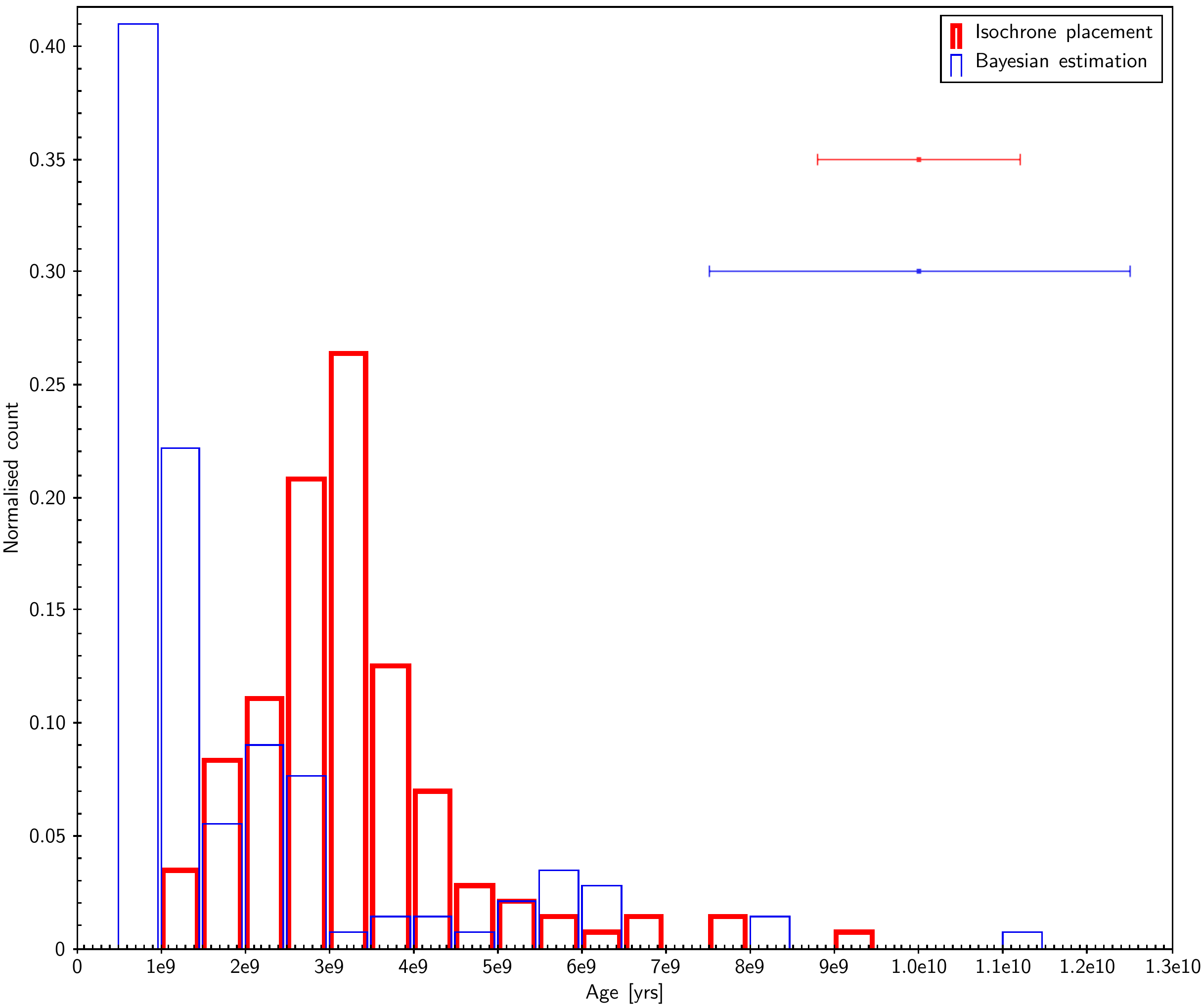}
 \caption{Isochrone placement vs. modal Bayesian ages. Age distribution of the perturbed synthetic stars with $4<M_V<8$.}
 \label{fig:DistrEta3Pert}
\end{figure}

Finally, we introduced a random Gaussian perturbation in the input $M_V$, $B-V$ and $\log{g}$ of the synthetic stars, considering Gaussian distributions with $3\sigma=1\%$ of the unperturbed values. We show the age distributions deriving from the isochrone placement and the modal Bayesian age in Fig. \ref{fig:DistrEta3Pert}. As expected, the distributions are broader and the mean error bars of the output results are higher. The mean age value of the isochrone placement does not change. The behaviour of the two techniques is similar to what we have just said about the ages derived from unperturbed input values.

In conclusion, the isochrone placement technique is the method chosen to compute the ages of the SWP.

\subsection{Stars with planets ages}
\begin{figure}
 \centering
 \includegraphics[width=\columnwidth]{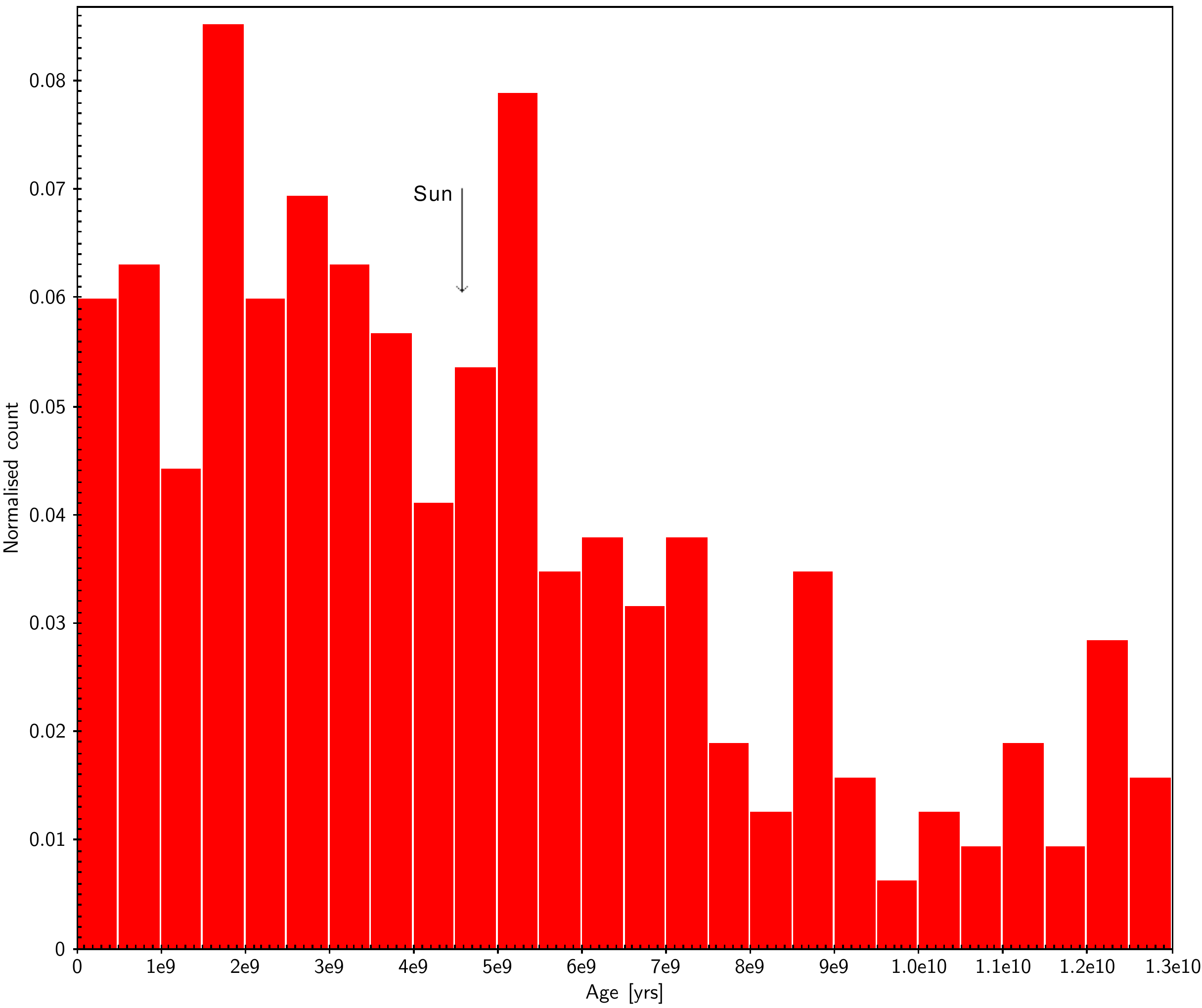}%
 \caption{Isochrone placement. SWP age distribution (317 stars).}
 \label{fig:SWP.IstoGauss}
\end{figure}

We analysed all the 326 stars belonging to the SWP Catalogue. Only nine of them have been removed because their observational parameters were not consistent with the theoretical ones. The age distribution of the remaining 317 stars determined using the isochrone placement technique is presented in Fig. \ref{fig:SWP.IstoGauss}. About 6\% of the stars are younger than 0.5 Gyr, and then the distribution reaches a peak at [1.5, 2) Gyr and after that it generally decreases. There is a non-negligible number of stars ($\sim7\%$) older than 11 Gyr. All the parameters of these stars are presented in Table \ref{tab:SWP}.

\begin{figure}
 \centering
 \includegraphics[width=\columnwidth]{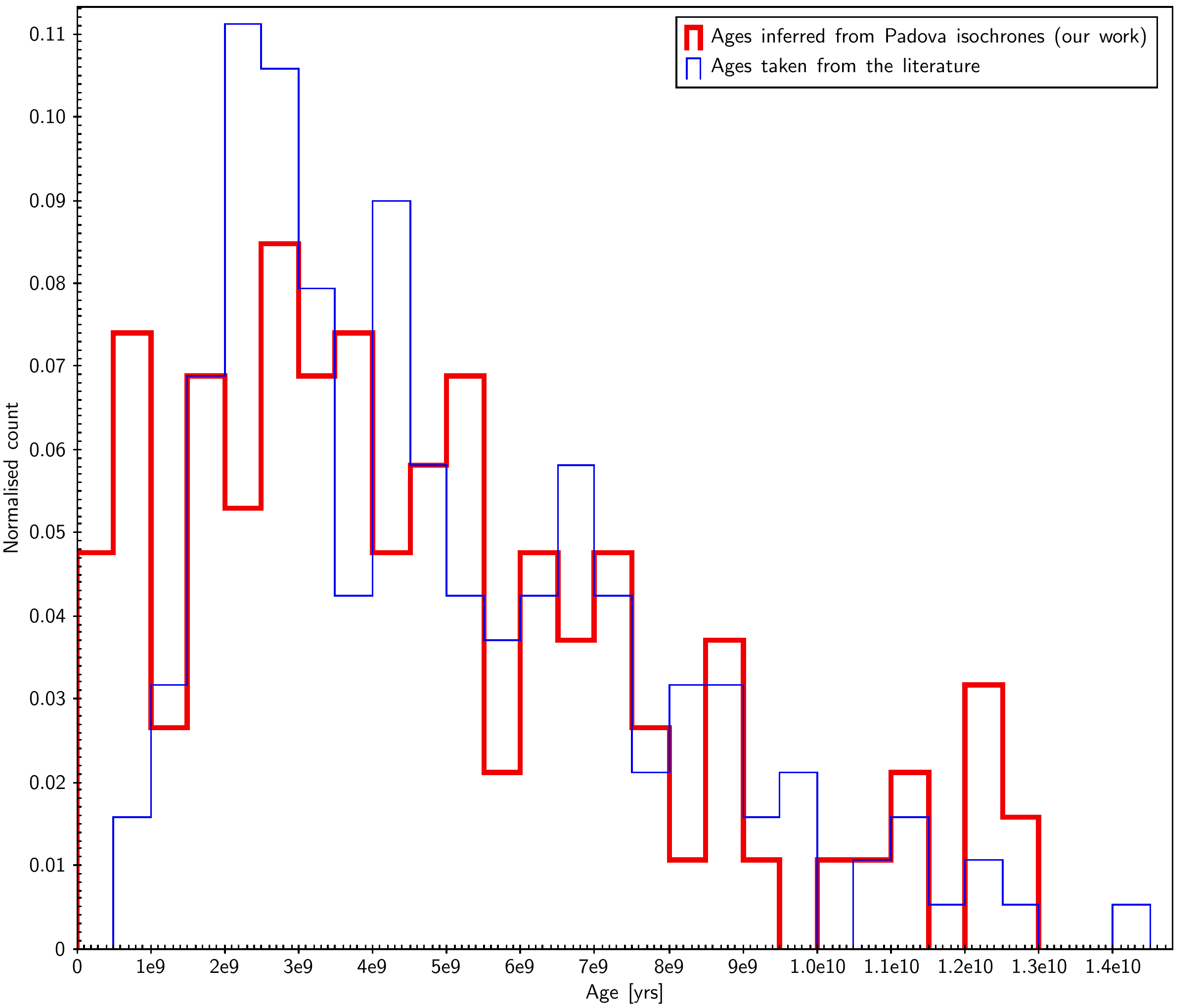}
 \caption{Comparison between the SWP age distribution derived here and coming from the literature (189 stars).}
 \label{fig:CfrLett.IstoSCP}
\end{figure}
\begin{figure}
 \centering
 \includegraphics[width=\columnwidth]{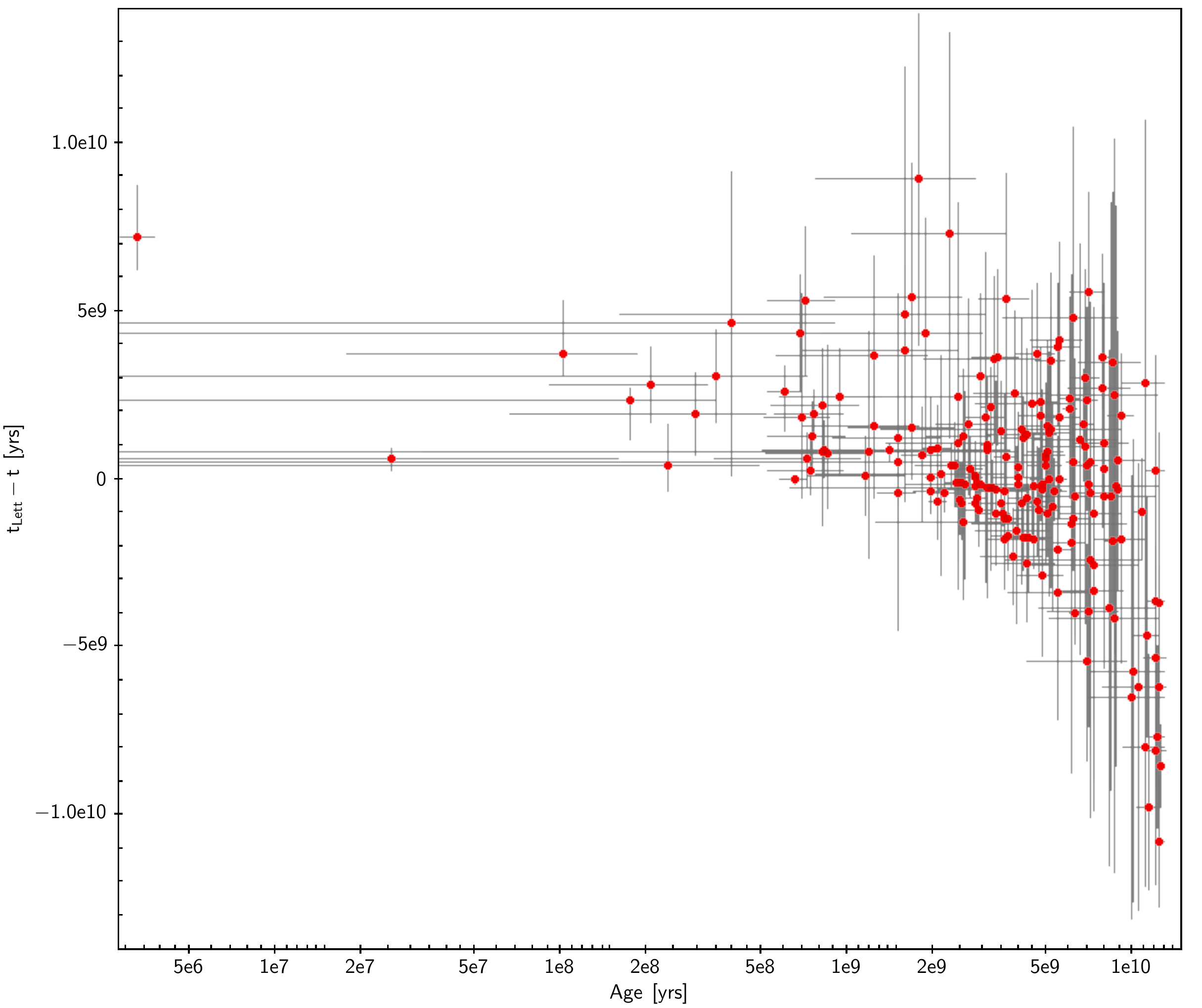}
 \caption{$t_{\mathrm{Lett}}-t$ vs. $t$ for the 189 SWP found in the literature.}
 \label{fig:CfrLett.SCP}
\end{figure}

The histogram of Fig. \ref{fig:SWP.IstoGauss} shows that there are SWP with all possible ages, with a preponderance of stars with ages < $\sim6$ Gyr. The median value ($\sim4$ Gyr) appears slightly lower than the age of the Sun. Of course, older MS stars are fainter (making it more difficult to identify planets). It is not obvious that there is a selection bias for the ages in Fig. \ref{fig:SWP.IstoGauss}. Very old or very young stars can have large uncertainties in age because the oldest isochrones overlap the pre-MS region. The frequency of stars in the [0, 0.5) Gyr bin is still an open question. The majority of the youngest SWP are hot Jupiter (HJs) hosts, and they tend to have high rotational velocity, so the check of activity did not discard the youngest isochrones. Actually, it might happen that such high rotational velocity does not reflect the stellar activity, while it can be due to the spin-up induced by the HJ, as suggested for example by \cite{poppenhaeger14}. Without definite indications from 
activity 
indices, another way to disentangle pre-MS from post-MS isochrones is to consider the stellar density $\rho_{\star}$, which is observationally available in the case of SWP detected through the transit method (see e.g. \cite{sozzetti07}). Here, $\rho_{\star}$ gives indications of the evolutionary stage of a star, and it has been used, for instance, by \cite{rouan12} to reject an extremely low age for CoRoT-23. We are planning of inserting this kind of check in the future development of our algorithm.

The literature reports the ages of 189 SWP, estimated using isochrones, but with different theoretical models and techniques. The superimposition of the consequent literature age distribution on what is found here is shown in Fig. \ref{fig:CfrLett.IstoSCP}. The two distributions are quite in agreement for ages older than $\sim4.5$ Gyr, while we found more stars in the domain of younger ages. In particular, in our sample, $\sim5\%$ of the stars has an age between 0 and 0.5 Gyr, while no stars in the literature fall in this first bin.

The difference $t_{\mathrm{Lett}}-t$ between the ages found in the literature and the values computed here is represented in Fig. \ref{fig:CfrLett.SCP}, which shows a wide spread in the age values. We provide new ages, which have the advantage of being derived using the same method and the same set of isochrones, and therefore useful for statistical investigations. Our age distribution in Fig. \ref{fig:CfrLett.IstoSCP} is broader than in the literature. Our data show an overabundance of young stars. Some young stars in the literature are judged to be older by our technique. This may happen for stars located on the red side of the isochrone interval, where the locus of the old isochrones is very near to the pre-MS one.

\begin{figure}
 \centering
 \includegraphics[width=\columnwidth]{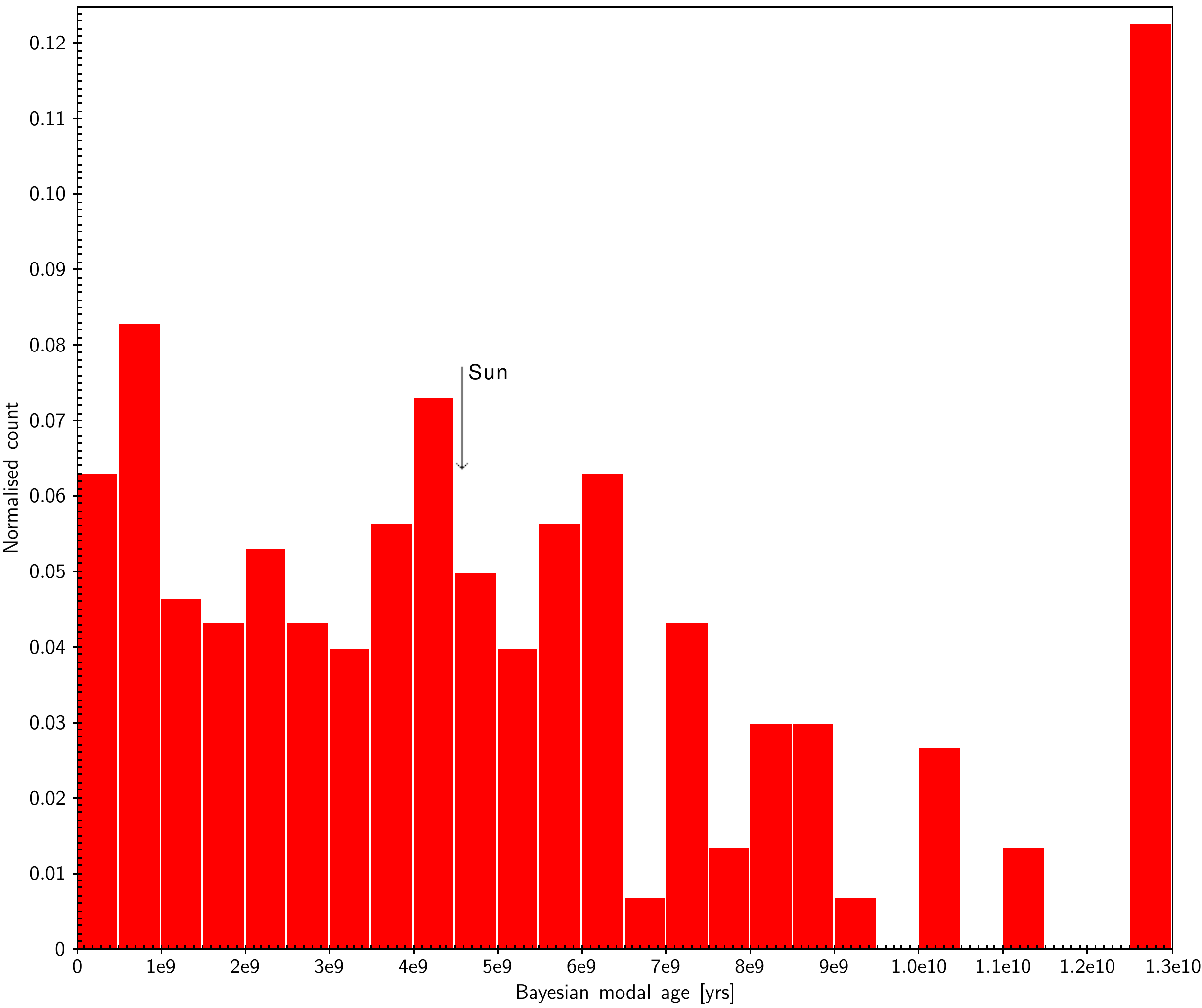}
 \caption{Bayesian estimation. SWP modal age distribution (302 stars).}
 \label{fig:SWP.IstoBayes}
\end{figure}

\begin{figure}
 \centering
 \includegraphics[width=\columnwidth]{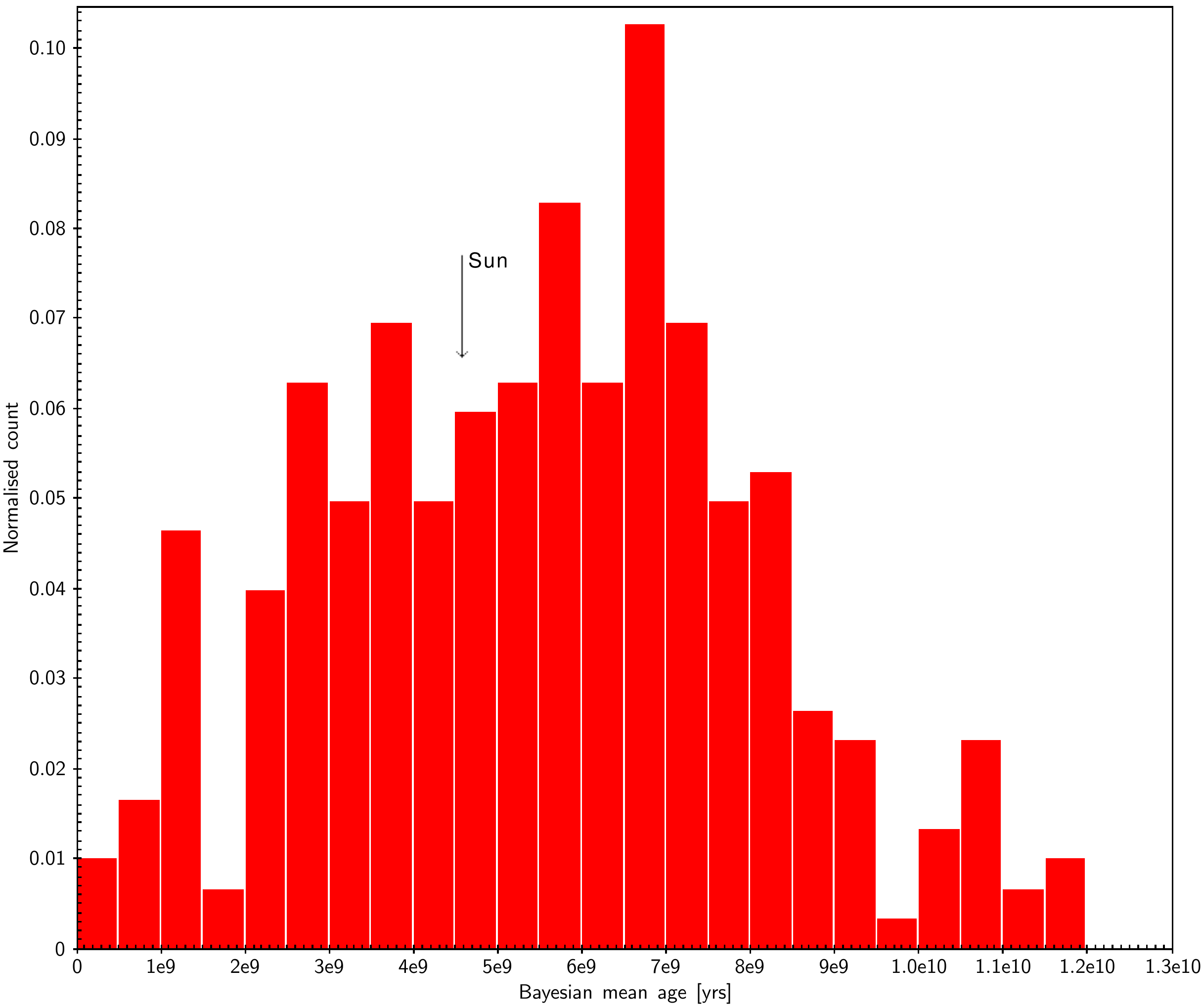}
 \caption{Bayesian estimation. SWP mean age distribution (302 stars).}
 \label{fig:SWP.IstoBayesM}
\end{figure}

Just for completeness, we also show the age distributions of 302 stars belonging to the SWP catalogue from the modal Bayesian estimation. Given that we find high frequencies in the first and in the last age bin of Fig. \ref{fig:SWP.IstoBayes}, it is again clear that the mode always tends to assign extreme age values to the stars. Figure \ref{fig:SWP.IstoBayesM} again suggests that the mean always produces a distribution centred on the middle of the age range. We avoid reporting the age distribution of SWP inferred from the median Bayesian age value, because it is very similar to Fig. \ref{fig:SWP.IstoBayesM} where the mean is employed as reference statistical index.

\begin{figure}
 \centering
 \includegraphics[width=\columnwidth]{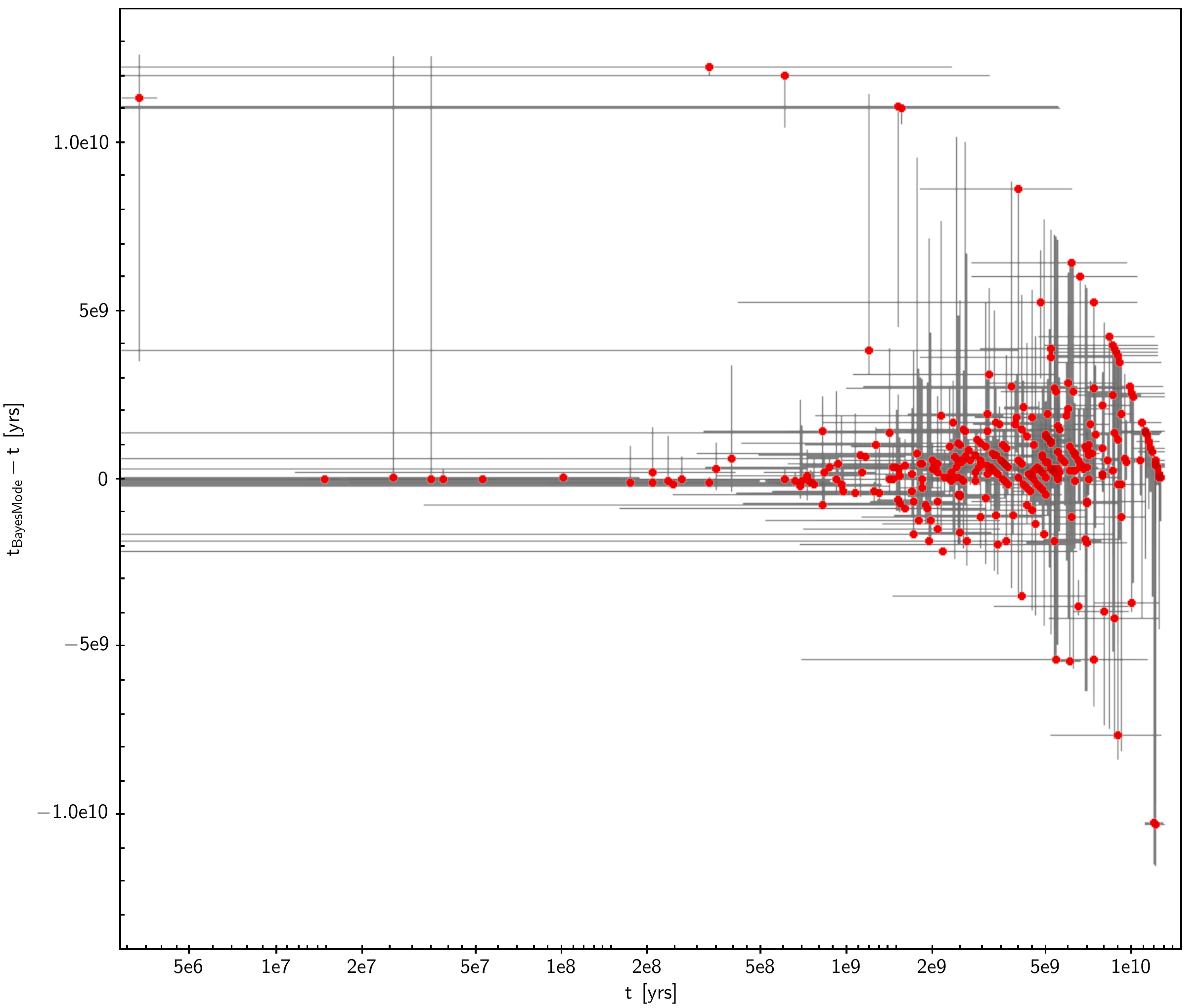}
 \caption{$t_{\mathrm{BayesMode}}-t$ vs. $t$ (302 stars).}
 \label{fig:CfrSCPGaussBayes}
\end{figure}

Figure \ref{fig:CfrSCPGaussBayes} shows the difference $t_{\mathrm{BayesMode}}-t$ between the modal Bayesian age $t_{\mathrm{BayesMode}}$ and the age $t$ computed through the isochrone placement represented versus $t$. Apart from the effect of the mode of attributing the most extreme age values available in the isochrone grids (visible from the locus of points in the upper right part of the figure), the other Bayesian-estimated age values appear slightly biased towards older ages. The median of the isochrone placement age distribution is $\sim4$ Gyr and the mean error is $\sim1.15$ Gyr, while the Bayesian modal age distribution has a median of $\sim4.25$ Gyr, and the mean error is $\sim1.75$ Gyr. This behaviour of the method contrasts with what is reported by \cite{haywood13}, who used both a $\chi^2$ minimization (somewhat comparable to our isochrone placement technique, even if they considered the $\chi^2$ minimization simply on the ($T_{\mathrm{eff}}$, $M_V$) plane) and the Bayesian technique described by 
\cite{jorgensen05} to compute the ages of nearby field stars. In their paper, the Bayesian technique does give younger ages, so we again emphasize the crucial role played by the details within the specific method implemented to compute the ages of field stars.

In the upper left-hand side of Figures \ref{fig:CfrLett.SCP} and \ref{fig:CfrSCPGaussBayes} there is a single isolated point, which is the star CoRoT-23. The paper that discusses its detection (\cite{rouan12}) reports $V_{\star}=15.63$ mag and $B_{\star}=16.96$ mag. $V_{\star}$ also agrees with the magnitude reported by \url{exoplanets.eu}\footnote{%
\url{http://exoplanets.eu}: The Extrasolar Planets Encyclopaedia.}. Instead, \url{exoplanets.org} reports $V_{\star}=16.96$ mag. Since \url{exoplanets.org} reports $B-V_{\star}=1.33$ (same value as inferred from \cite{rouan12}), it is likely that $B_{\star}=16.96\neq V_{\star}$. However, after applying the isochrone placement with the new photometry, our code does not converge on any age value. This new photometry, in fact, moves the star farther out from the set of isochrones. This means that, first of all, there is a problem in the source of the photometry. Moreover, CoRoT-23 is a peculiar system considering its age and the orbital eccentricity of the hosted planet (see \cite{rouan12}). As already said, taking the observational $\rho_{\star}$ into account may help give a better answer to this problem.

Either way, using $V_{\star}=16.96$ mag, we obtained an isochrone placement age $t_{\star\mathrm{,Isoc}}=3.3$ Myr, while the Bayesian estimation gives $t_{\star\mathrm{,Bayes}}=11.3$ Gyr. This is why CoRoT-23 appears as an outlier in Fig. \ref{fig:CfrSCPGaussBayes}. Thus $t_{\star\mathrm{,Bayes}}\gg t_{\star\mathrm{,Isoc}}$ because, even if the Bayesian pdf had a peak corresponding to a very low age, such a peak was judged as a spike by the smoothing through the polynomial interpolation, so it was erased. The next major peak was then at 11.3 Gyr.

\begin{figure}
 \centering
 \includegraphics[width=\columnwidth]{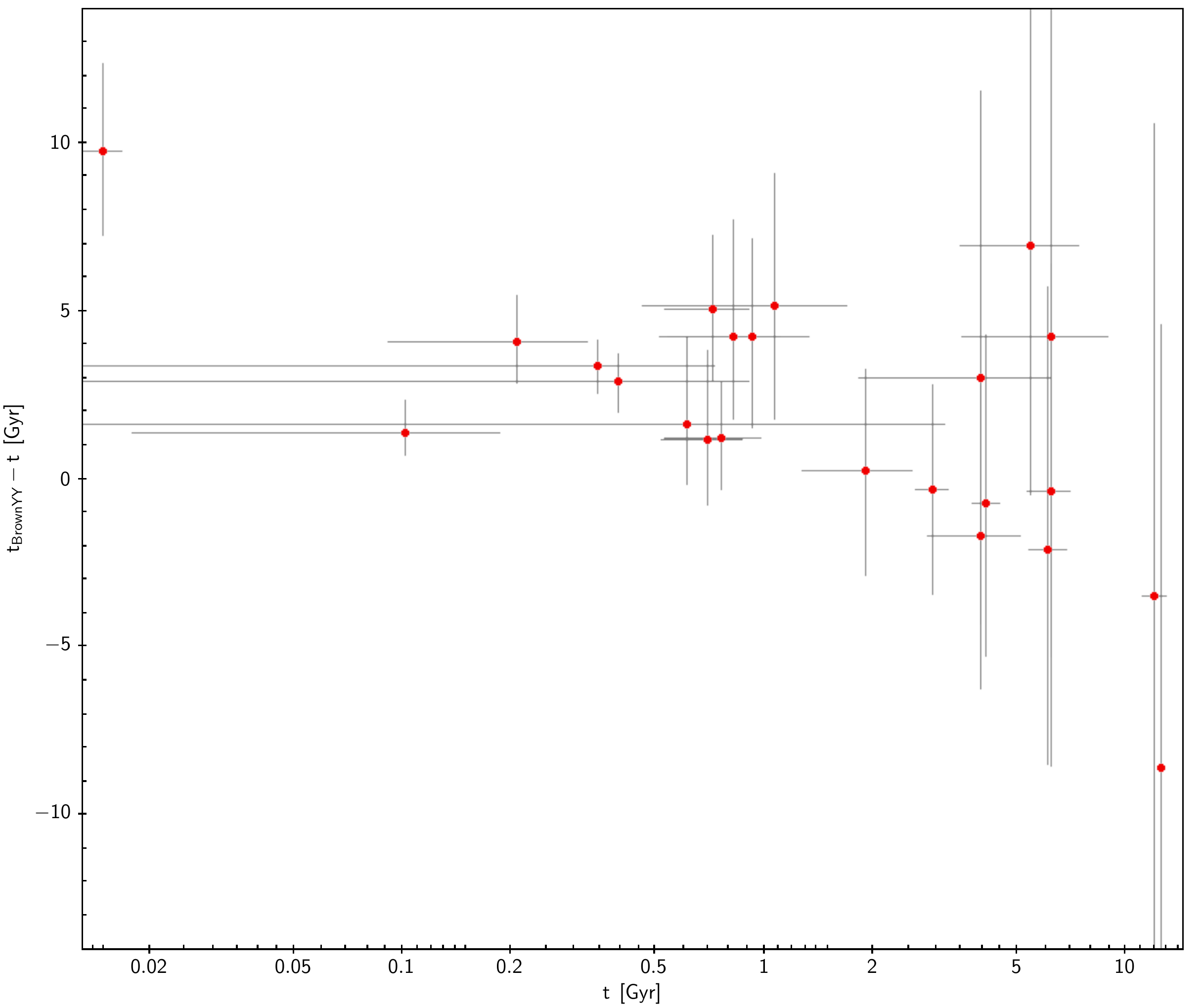}
 \caption{$t_{\mathrm{BrownYY}}-t$ vs. $t$ for the 24 SWP in common with \cite{brown14}.}
 \label{fig:BrownYYvsPD}
\end{figure}

Finally, we considered the work of \cite{brown14}, who used different isochrones and gyrochronological relations to assess the ages of a sample of SWP. We found 24 SWP in common with \cite{brown14}. The difference $t_{\mathrm{BrownYY}}-t$ between the ages $t_{\mathrm{BrownYY}}$ computed by Brown using the YY isochrones (\cite{demarque04}) and the ages $t$ computed here through the isochrone placement are represented versus $t$ in Fig. \ref{fig:BrownYYvsPD}. Considering the small sample of stars, the spread is consistent with the uncertainties of the ages of MS stars. The outlier in the upper left-hand side in Fig. \ref{fig:BrownYYvsPD} is WASP-2. It is in the pre-MS region of the CMD, but we do not have any activity index, which could allow us to discard the youngest isochrones.

\begin{figure}
 \centering
 \includegraphics[width=\columnwidth]{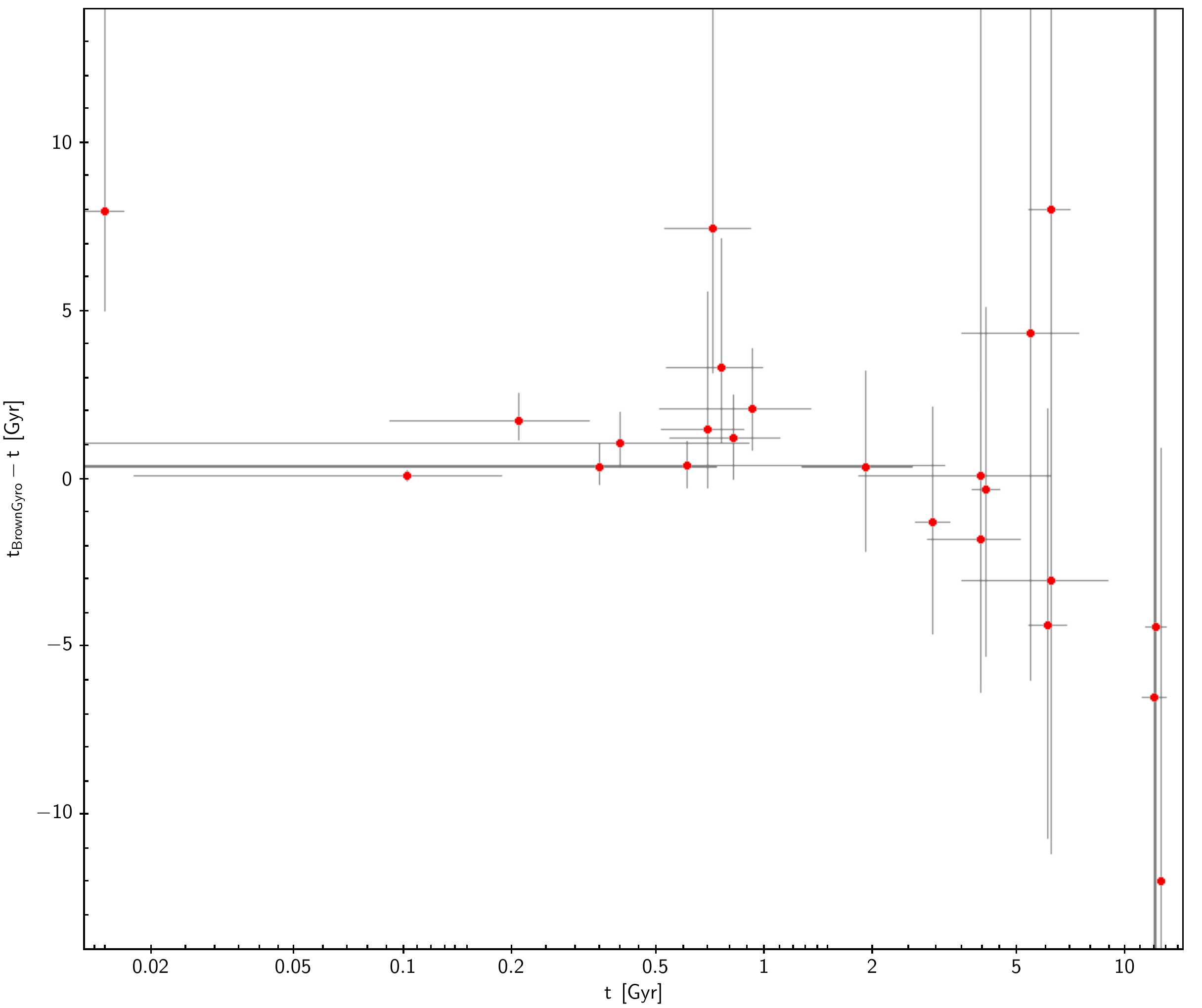}
 \caption{$t_{\mathrm{BrownGyro}}-t$ vs. $t$ for the 24 SWP in common with \cite{brown14}.}
 \label{fig:GyrovsPD}
\end{figure}

As stated by \cite{brown14} in his paper, isochrones tend to give ages older than gyrochronology. If, instead, we employ our determination of the SWP ages $t$ through the isochrone placement technique, we find that isochrones can also give ages younger than gyrochronology as clarified by Fig. \ref{fig:GyrovsPD}, where the gyrochronological ages $t_{\mathrm{BrownGyro}}$ obtained from the relation of \cite{barnes10} are used. Considering the typical age uncertainties, the agreement with Brown's gyrochronological ages is very good with a lower dispersion if compared with the ages derived from YY models. It is likely that those stars, which turn out to be the oldest from the isochrone placement, do not have an accurate gyrochronological age considering that the age from the rotational velocity to age is very uncertain after some billion years.


\subsection{Impact of the input parameters on the output ones}
In this section we briefly discuss the sensitivity of the output parameters derived through the isochrone placement technique to the parameters assumed as input. We simulated the Sun, adopting the input parameters listed in the input column of Table \ref{tab:SunParameters}. These are the same parameters as the isochrones we used to fit of the Sun. We attributed the typical uncertainties of the SWP catalogue stars to them, i.e. $\Delta\log{g}=0.1$ dex, $\Delta L$ comes from the error propagation of $\Delta m_{\mathrm{bol}}=0.03$ mag and $\bar{\Delta}d\sim7\%$; we have $\Delta L\sim17\%$, $\Delta R\sim10\%$, and $\Delta M\sim40\%$.
\begin{table}
 \caption{\emph{Sun} parameters}
 \label{tab:SunParameters}
 \begin{tabular}{lcc}
  \hline
  Parameters		&	Input		&	Output		\\
  \hline
  [Fe/H]		&	0		&			\\
  $V$			&	$-26.739$	&	 		\\
  $d$ [AU]		&	$1\pm7\%$	&			\\
  $B-V$			&	0.667		&			\\
  $\log{g}$ [cgs]	&	$4.432\pm0.1$	&	$4.432\pm0.002$	\\
  $T_{\mathrm{eff}}$ [K]&	$5778\pm1\%$	&	$5778\pm3$	\\
  $L$ [$L_{\odot}$]	&	$1.00\pm0.17$	&	$1.0001\pm0.0005$\\
  $R$ [$R_{\odot}$]	&	$1.00\pm0.10$	&	$1.000\pm0.001$	\\
  $M$ [$M_{\odot}$]	&	$1.00\pm0.44$	&	$1.003\pm0.002$	\\
  $t$ [Gyr]		&			&	$4.5\pm0.1$	\\
 \hline
 \end{tabular}
\end{table}
The output parameters obtained according to the isochrones are shown in the output column of Table \ref{tab:SunParameters}. They are consistent with the expected solar parameters, and among them, the age is correct and characterized by very high precision.

To evaluate how changes in both the photometry and the spectroscopy propagate into the results, we perturbed the input $V$ and $B-V$ of the Sun alternatively by $\pm0.01$ mag and $\log{g}$ by $\pm0.01$ dex. Table \ref{tab:SunPert} presents what we obtained, also listing the percentual variations $\Delta_{\mathrm{oo}}$ of the output data assuming as reference values the output values obtained without any perturbation of the input parameters.

\begin{table*}
 \centering
 \caption{Output parameters after the perturbation of the \emph{Sun}.}
 \label{tab:SunPert}
 \begin{tabular}{*{7}{c}}
  \hline
  \textit{Sun}		&	$t$ [Gyr]	&	$T_{\mathrm{eff}}$ [K]	&	$L$	[$L_{\odot}$]	& $R$ [$R_{\odot}$]	      &	$M$ [$M_{\odot}$]    &	$\log g$ [cgs]	\\	
  \hline
  Reference data	&	$4.5\pm0.1$	&	5778			&	1.0001			&	1.000			      &	      1.003	      &	      4.432		\\
  \hline
  $\Delta V=+0.01$	&	$4.2\pm0.9$	&	5780			&	0.991			&	0.995			      &	      1.005	      &	      4.438		\\
  $\Delta_{\mathrm{oo}}$&	$-6.7\%$	&	+0.03\%			&	$-0.9\%$		&    $-0.5\%$			      &	      +0.2\%	      &	      +0.1\%		\\
  \hline
  $\Delta V=-0.01$	&	$4.5\pm0.7$	&	5783			&	1.008			&	1.002			      &	      1.004	      &	     4.431		\\
  $\Delta_{\mathrm{oo}}$&	0\%		&     +0.09\%			&	+0.8\%			&      +0.2\%			      &	      +0.1\%	      &	     $-0.02\%$		\\
  \hline
  $\Delta(B-V)=+0.01$	&	$5.3\pm1.2$	&	5756			&	1.002			&	1.009			      &	      0.991	      &      4.420		\\
  $\Delta_{\mathrm{oo}}$&	+17.8\%		&	$-0.4\%$		&	+0.2\%			&      +0.9\%			      &	      $-1.2\%$	      &	     $-0.3\%$		\\
  \hline
  $\Delta(B-V)=-0.01$	&	$2.7\pm1.3$	&	5819			&	0.995			&	0.983			      &	      1.028	      &	     4.458		\\
  $\Delta_{\mathrm{oo}}$&      $-40\%$		&	+0.7\%			&	$-0.5\%$		&    $-1.7\%$			      &	      +2.5\%	      &	      +0.6\%		\\
  \hline
  $\Delta\log{g}=+0.01$ &	$4.2\pm0.9$	&	5783			&	0.999			&	0.998			       &     1.006	      &	      4.436		\\
  $\Delta_{\mathrm{oo}}$&	$-6.7\%$	&	+0.09\%			&	$-0.1\%$		&    $-0.2\%$			      &	      +0.3\%	      &	      +0.09\%		\\
  \hline
  $\Delta\log{g}=-0.01$ &	$4.3\pm1.0$	&	5781			&	1.000			&	0.999			       &     1.005	      &       4.434		\\
  $\Delta_{\mathrm{oo}}$&	$-4.4\%$	&	+0.05\%			&	$-0.01\%$		&    $-0.1\%$			      &	      +0.2\%	      &	      +0.05\%		\\
  \hline
\end{tabular}
\end{table*}

Table \ref{tab:SunPert} shows that a variation of only 0.01 mag in $V$ or $B-V$ can lead to variations in the mean output values up to $\sim40\%$ in the age and up to $\sim2\%$ in the other parameters. This level of sensitivity for the derived age to the input photometry is understandable considering that we chose a MS star to perform these tests. In this region of the HRD, the isochrones are very close, so that even a small variation in magnitude and colours completely moves onto an isochrone that corresponds to a different age value. It is reasonable that if the star were in a different region of the HRD, where the isochrones are not so closed, the change in age induced by the perturbation of the input photometry would not be evident like this. Anyway, a very precise photometry is required in order to attribute the correct age to a star. On the other hand, even a slight perturbation in the input $\log{g}$ by only 0.01 dex induces a variation in the ages up to $\sim7\%$, while the variations in the 
other parameters can be considered almost negligible.

A final observation concerns the uncertainties accompanying the age values reported in Table \ref{tab:SunPert}, which also turn out to be an order of magnitude greater than the uncertainty of the output age of the unperturbed Sun. This is because the perturbations trigger so-called artificial stars, and the big uncertainties reflect the difficulty of properly matching all the input parameters in the isochrone grids, where parameters referring to stars that are supposed to exist are tabulated. On the other hand, since the physics underlying the isochrones is well performed and the Sun does exist, entering with consistent input stellar properties gives high precision output.

\section{Conclusions}\label{sec:conclusion}

In this paper we uniformly derived the ages of 317 planet-hosting stars. We checked the reliability and accuracy of two techniques (isochrone placement and Bayesian estimation) from a sample of 3.2 Gyr synthetic stars. We found that the isochrone placement technique produces the expected age value. Instead, the estimation of age through the Bayesian statistics (using an explicit probability density function) suffers the problem of primarily selecting the extreme age values available in the isochrones grid if we adopt the modal value of the pdf. It is worth emphasizing that this bias is present, although we used stars with well-defined ages. The mean or the median values of the pdf produce a bias towards the middle of the available age range and the age dispersion is large.

We found that $\sim6\%$ of the stars with planets are younger than 0.5 Gyr: the reliability of this frequency will be subject to further investigations. The age distribution shows a peak in correspondence of the bin [1.5, 2) Gyr, then it decreases and $\sim7\%$ of the stars are older than 11 Gyr. Approximately $60\%$ of the stars in this sample are younger than 5 Gyr. 

We found that a perturbation of the input $V$ or $B-V$ by only 0.01 mag can lead to variations up to 40\% in the estimated age of a MS Sun-like star. These are the typical uncertainties that characterize the stars in our sample. Instead, a perturbation in the input $\log{g}$ by 0.01 dex can lead to variations up to 7\% in the estimated age. The final accuracy depends on the combination of the input errors. Important uncertainties and biases can also be produced by the adopted stellar models, in particular by treating the element diffusion. We found that one of the most important consequences of the element diffusion is the choice of the correct isochrone metallicity because the initially metallicity can be considerably higher than the metallicity of the stellar atmosphere after some billion years for solar-type stars. If this effect is ignored, the ages can be overestimated up to about $25\%$.

Finally, we confirmed the presence of some stars with planets located on the left-hand side of the solar main sequence (as also noted by \cite{brown14}) and suggested further photometric observations to identify the source of the problem. Once the observational data is confirmed, we will investigate the nature using models that implement rotation. In fact, in the typical range of luminosities of MS stars with planets, MS isochrones that consider rotation are bluer than models that do not take rotation into account, as can be seen in \cite{yang13}.

\begin{acknowledgements}
 We thank the anonimous referee for the useful, thoughtful and deep comments that helped us in improving our paper.
 
 V.~N.~acknowledges partial support from INAF-OAPd through the grant ``Analysis of HARPS-N data in the framework of GAPS project'' (\#19/2013) and ``Studio preparatorio per le osservazioni della missione ESA/CHEOPS'' (\#42/2013).
 
 This research has made use of the Exoplanet Orbit Database and the Exoplanet Data Explorer at exoplanets.org.
\end{acknowledgements}

\onllongtab{
%
\begin{longtab}
\scriptsize
\begin{longtable}{lrrrrrrrrrrrr}
\caption{\label{tab:SWP} SWP parameters determined through Padova Isochrones.}\\
\hline\hline
Star & $t$ & $\Delta t$ & $T_{\mathrm{eff}}$ & $\Delta T_{\mathrm{eff}}$ & $L$ & $\Delta L$ & $M$ & $\Delta M$ & $\log{g}$ & $\Delta\log{g}$ & $R$ & $\Delta R$ \\
     & (Gyr) & (Gyr) & (K) & (K) & ($L_{\odot}$) & ($L_{\odot}$) & ($M_{\odot}$) & ($M_{\odot}$) & (cm/s$^2$) & (cm/s$^2$) & ($R_{\odot}$) & ($R_{\odot}$) \\
\hline
\endfirsthead
\caption{continued.}\\
\hline\hline
Star & $t$ & $\Delta t$ & $T_{\mathrm{eff}}$ & $\Delta T_{\mathrm{eff}}$ & $L$ & $\Delta L$ & $M$ & $\Delta M$ & $\log{g}$ & $\Delta\log{g}$ & $R$ & $\Delta R$ \\
     & (Gyr) & (Gyr) & (K) & (K) & ($L_{\odot}$) & ($L_{\odot}$) & ($M_{\odot}$) & ($M_{\odot}$) & (cm/s$^2$) & (cm/s$^2$) & ($R_{\odot}$) & ($R_{\odot}$) \\
\hline
\endhead
\hline
\endfoot
  WASP-26 & 0.7 & 0.2 & 6245 & 7 & 1.603 & 0.001 & 1.15 & 0.005 & 4.422 & 0.004 & 1.084 & 0.003\\
  HD 1502 & 2.5 & 0.1 & 5059 & 13 & 9.75 & 0.03 & 1.59 & 0.01 & 3.41 & 0.01 & 4.07 & 0.03\\
  HD 2039 & 4.4 & 0.8 & 5935 & 64 & 2.18 & 0.02 & 1.2 & 0.03 & 4.22 & 0.03 & 1.4 & 0.04\\
  HIP 2247 & 10.5 & 2.6 & 4692 & 9 & 0.212 & 0.001 & 0.72 & 0.01 & 4.6 & 0.01 & 0.7 & 0.01\\
  HD 2638 & 1.9 & 2.6 & 5160 & 24 & 0.407 & 0.004 & 0.89 & 0.02 & 4.58 & 0.02 & 0.8 & 0.01\\
  HAT-P-16 & 0.8 & 0.2 & 6326 & 12 & 2.125 & 0.002 & 1.25 & 0.01 & 4.36 & 0.01 & 1.22 & 0.01\\
  HD 3651 & 4.6 & 2.9 & 5284 & 29 & 0.51 & 0.005 & 0.91 & 0.02 & 4.53 & 0.02 & 0.85 & 0.01\\
  HD 4208 & 6.6 & 2.1 & 5717 & 33 & 0.71 & 0.004 & 0.86 & 0.02 & 4.5 & 0.03 & 0.86 & 0.01\\
  HD 4308 & 1.6 & 4.0 & 5714 & 61 & 1.03 & 0.01 & 0.95 & 0.05 & 4.38 & 0.02 & 1.04 & 0.03\\
  HD 4203 & 6.3 & 1.0 & 5666 & 43 & 1.68 & 0.01 & 1.12 & 0.03 & 4.22 & 0.03 & 1.35 & 0.03\\
  HD 4313 & 2.0 & 0.1 & 5006 & 25 & 13.9 & 0.1 & 1.72 & 0.03 & 3.29 & 0.03 & 4.9 & 0.1\\
  HD 4732 & 2.3 & 0.2 & 4994 & 32 & 14.8 & 0.2 & 1.61 & 0.05 & 3.22 & 0.03 & 5.1 & 0.1\\
  HD 5319 & 3.6 & 0.4 & 4941 & 40 & 8.2 & 0.1 & 1.4 & 0.1 & 3.4 & 0.04 & 3.9 & 0.1\\
  HD 5388 & 5.5 & 0.5 & 6195 & 78 & 4.43 & 0.02 & 1.11 & 0.03 & 3.98 & 0.04 & 1.8 & 0.1\\
  HD 5891 & 1.5 & 0.1 & 4915 & 23 & 32.9 & 0.2 & 1.87 & 0.04 & 2.9 & 0.02 & 7.9 & 0.1\\
  HD 6434 & 12.2 & 0.8 & 5907 & 71 & 1.208 & 0.004 & 0.83 & 0.03 & 4.31 & 0.01 & 1.029 & 0.004\\
  HIP 5158 & 4.9 & 3.7 & 4592 & 11 & 0.185 & 0.001 & 0.74 & 0.01 & 4.63 & 0.02 & 0.68 & 0.01\\
  HD 6718 & 6.0 & 2.4 & 5818 & 57 & 1.07 & 0.01 & 0.98 & 0.04 & 4.42 & 0.04 & 1.01 & 0.02\\
  HD 7199 & 9.5 & 2.4 & 5349 & 38 & 0.7 & 0.01 & 0.93 & 0.02 & 4.42 & 0.03 & 0.98 & 0.02\\
  HD 7449 & 2.4 & 1.5 & 6070 & 45 & 1.24 & 0.01 & 1.04 & 0.03 & 4.45 & 0.02 & 1.01 & 0.01\\
  HD 7924 & 3.0 & 1.8 & 5216 & 13 & 0.364 & 0.001 & 0.81 & 0.01 & 4.6 & 0.01 & 0.74 & 0.01\\
  HD 8535 & 2.1 & 0.9 & 6200 & 50 & 1.85 & 0.01 & 1.17 & 0.02 & 4.36 & 0.02 & 1.18 & 0.02\\
  HD 8574 & 5.0 & 0.1 & 6065 & 6 & 2.335 & 0.001 & 1.144 & 0.003 & 4.21 & 0.03 & 1.39 & 0.01\\
  HD 9446 & 2.0 & 1.5 & 5771 & 28 & 0.924 & 0.005 & 1.03 & 0.02 & 4.48 & 0.02 & 0.96 & 0.01\\
  WASP-18 & 0.2 & 0.3 & 6188 & 10 & 1.442 & 0.002 & 1.14 & 0.01 & 4.45 & 0.01 & 1.047 & 0.004\\
  HD 10180 & 3.9 & 1.1 & 5962 & 47 & 1.48 & 0.01 & 1.09 & 0.02 & 4.35 & 0.04 & 1.14 & 0.03\\
  HD 10647 & 3.2 & 1.2 & 6128 & 41 & 1.57 & 0.01 & 1.08 & 0.03 & 4.39 & 0.03 & 1.1 & 0.02\\
  HD 10697 & 7.1 & 0.1 & 5715 & 26 & 2.84 & 0.01 & 1.129 & 0.005 & 4.01 & 0.02 & 1.72 & 0.04\\
  HD 11506 & 1.6 & 0.9 & 5833 & 28 & 1.17 & 0.01 & 1.12 & 0.02 & 4.43 & 0.02 & 1.06 & 0.01\\
  HD 11977 & 1.2 & 0.4 & 5001 & 32 & 60.1 & 0.5 & 2.0 & 0.2 & 2.7 & 0.1 & 10.4 & 0.2\\
  HAT-P-32 & 0.1 & 0.1 & 6594 & 6 & 2.251 & 0.001 & 1.221 & 0.004 & 4.395 & 0.003 & 1.152 & 0.002\\
  HD 12661 & 1.8 & 0.5 & 5765 & 14 & 1.104 & 0.003 & 1.11 & 0.01 & 4.43 & 0.01 & 1.05 & 0.01\\
  alpha Ari & 3.4 & 1.9 & 4563 & 26 & 79.2 & 1.0 & 1.4 & 0.2 & 2.3 & 0.1 & 14.3 & 0.2\\
  HAT-P-29 & 0.3 & 0.2 & 6137 & 7 & 1.6 & 0.001 & 1.2 & 0.01 & 4.412 & 0.004 & 1.121 & 0.003\\
  HD 13931 & 6.8 & 0.6 & 5868 & 24 & 1.49 & 0.01 & 1.04 & 0.01 & 4.3 & 0.03 & 1.18 & 0.02\\
  WASP-33 & 0.04 & 0.01 & 7268 & 3 & 5.23 & 0.005 & 1.539 & 0.001 & 4.298 & 0.001 & 1.445 & 0.002\\
  HD 16175 & 3.2 & 0.2 & 6048 & 35 & 3.3 & 0.01 & 1.34 & 0.01 & 4.12 & 0.03 & 1.66 & 0.04\\
  81 Cet & 2.5 & 0.9 & 4825 & 41 & 60.0 & 0.8 & 1.6 & 0.2 & 2.5 & 0.1 & 11.1 & 0.3\\
  HD 16760 & 1.3 & 0.9 & 5518 & 11 & 0.58 & 0.002 & 0.93 & 0.01 & 4.56 & 0.01 & 0.835 & 0.005\\
  iota Hor & 1.5 & 0.6 & 6148 & 31 & 1.68 & 0.01 & 1.17 & 0.01 & 4.38 & 0.02 & 1.14 & 0.01\\
  HD 17156 & 4.1 & 0.4 & 5997 & 27 & 2.5 & 0.01 & 1.23 & 0.01 & 4.19 & 0.02 & 1.47 & 0.02\\
  HD 18742 & 2.9 & 0.2 & 5009 & 22 & 11.9 & 0.1 & 1.48 & 0.03 & 3.28 & 0.02 & 4.6 & 0.1\\
  WASP-11 & 8.7 & 3.5 & 4884 & 16 & 0.28 & 0.002 & 0.77 & 0.02 & 4.58 & 0.02 & 0.74 & 0.01\\
  HIP 14810 & 8.7 & 2.0 & 5535 & 51 & 0.99 & 0.01 & 0.98 & 0.02 & 4.35 & 0.03 & 1.08 & 0.03\\
  HAT-P-25 & 11.2 & 1.9 & 5099 & 18 & 0.478 & 0.003 & 0.86 & 0.01 & 4.47 & 0.02 & 0.89 & 0.01\\
  HD 20794 & 11.2 & 1.8 & 5610 & 25 & 0.645 & 0.003 & 0.8 & 0.01 & 4.47 & 0.02 & 0.85 & 0.02\\
  HD 20782 & 5.6 & 1.2 & 5878 & 35 & 1.22 & 0.01 & 1.01 & 0.02 & 4.38 & 0.03 & 1.07 & 0.02\\
  HD 20868 & 8.4 & 3.7 & 4811 & 14 & 0.255 & 0.002 & 0.76 & 0.02 & 4.59 & 0.02 & 0.73 & 0.01\\
  WASP-22 & 0.2 & 0.1 & 6105 & 3 & 1.2498 & 5.0E-4 & 1.101 & 0.004 & 4.472 & 0.002 & 1.001 & 0.001\\
  epsilon Eri & 1.4 & 1.7 & 5100 & 16 & 0.335 & 0.002 & 0.83 & 0.01 & 4.61 & 0.01 & 0.74 & 0.01\\
  HD 23127 & 4.8 & 0.6 & 5843 & 52 & 3.01 & 0.03 & 1.26 & 0.04 & 4.06 & 0.02 & 1.71 & 0.03\\
  HD 23079 & 5.1 & 1.0 & 6003 & 36 & 1.372 & 0.005 & 1.01 & 0.02 & 4.37 & 0.04 & 1.08 & 0.02\\
  HD 22781 & 7.3 & 3.1 & 5170 & 18 & 0.322 & 0.002 & 0.74 & 0.02 & 4.6 & 0.02 & 0.71 & 0.01\\
  HD 23596 & 5.0 & 0.7 & 5953 & 48 & 2.63 & 0.03 & 1.2 & 0.04 & 4.14 & 0.03 & 1.53 & 0.04\\
  HD 24040 & 4.8 & 0.8 & 5917 & 52 & 1.81 & 0.01 & 1.14 & 0.02 & 4.27 & 0.02 & 1.28 & 0.03\\
  HD 25171 & 4.9 & 0.8 & 6125 & 51 & 1.92 & 0.01 & 1.08 & 0.02 & 4.28 & 0.04 & 1.23 & 0.03\\
  HD 27894 & 2.2 & 4.2 & 4898 & 55 & 0.37 & 0.01 & 0.89 & 0.03 & 4.52 & 0.03 & 0.85 & 0.03\\
  XO-3 & 2.0 & 1.0 & 6634 & 11 & 7.04 & 0.01 & 1.41 & 0.03 & 3.97 & 0.03 & 2.01 & 0.01\\
  HD 28254 & 7.9 & 0.2 & 5607 & 19 & 2.13 & 0.01 & 1.1 & 0.01 & 4.1 & 0.02 & 1.54 & 0.03\\
  HD 28185 & 4.8 & 4.4 & 5609 & 41 & 1.18 & 0.01 & 1.0 & 0.1 & 4.33 & 0.03 & 1.15 & 0.03\\
  epsilon Tau & 0.7 & 0.1 & 4878 & 8 & 81.9 & 0.4 & 2.6 & 0.1 & 2.65 & 0.01 & 12.7 & 0.1\\
  HD 28678 & 4.3 & 1.1 & 4828 & 46 & 21.3 & 0.3 & 1.3 & 0.1 & 2.9 & 0.1 & 6.6 & 0.2\\
  HD 30177 & 4.8 & 1.5 & 5607 & 47 & 1.04 & 0.01 & 1.06 & 0.02 & 4.38 & 0.03 & 1.09 & 0.02\\
  HD 30562 & 4.4 & 0.6 & 5983 & 37 & 2.82 & 0.01 & 1.25 & 0.03 & 4.14 & 0.02 & 1.57 & 0.03\\
  HD 30856 & 4.7 & 0.9 & 4952 & 31 & 8.5 & 0.1 & 1.3 & 0.1 & 3.34 & 0.04 & 4.0 & 0.1\\
  HD 33142 & 3.3 & 0.4 & 5005 & 41 & 9.0 & 0.1 & 1.4 & 0.1 & 3.39 & 0.03 & 4.0 & 0.1\\
  HD 33283 & 3.6 & 0.6 & 5985 & 57 & 4.37 & 0.02 & 1.39 & 0.04 & 3.99 & 0.03 & 1.95 & 0.04\\
  HD 32518 & 6.4 & 1.5 & 4599 & 41 & 46.4 & 0.9 & 1.2 & 0.1 & 2.4 & 0.1 & 10.8 & 0.3\\
  HD 33636 & 2.5 & 1.1 & 5979 & 28 & 1.08 & 0.003 & 1.01 & 0.02 & 4.46 & 0.02 & 0.97 & 0.01\\
  HD 290327 & 11.8 & 1.2 & 5525 & 20 & 0.747 & 0.004 & 0.86 & 0.01 & 4.41 & 0.01 & 0.95 & 0.02\\
  HD 37124 & 11.8 & 1.2 & 5763 & 22 & 0.839 & 0.003 & 0.81 & 0.01 & 4.41 & 0.01 & 0.92 & 0.02\\
  HD 39091 & 3.4 & 0.6 & 6013 & 18 & 1.532 & 0.004 & 1.11 & 0.01 & 4.35 & 0.01 & 1.15 & 0.01\\
  HD 37605 & 1.8 & 1.0 & 5380 & 13 & 0.602 & 0.002 & 0.98 & 0.01 & 4.52 & 0.01 & 0.89 & 0.01\\
  HD 38801 & 4.8 & 0.3 & 5338 & 59 & 3.7 & 0.1 & 1.28 & 0.02 & 3.83 & 0.02 & 2.3 & 0.1\\
  HD 40307 & 7.0 & 4.2 & 4948 & 19 & 0.24 & 0.002 & 0.7 & 0.02 & 4.63 & 0.02 & 0.67 & 0.01\\
  WASP-49 & 1.1 & 0.6 & 5809 & 11 & 0.73 & 0.002 & 0.94 & 0.01 & 4.55 & 0.01 & 0.845 & 0.004\\
  HD 43197 & 3.1 & 2.0 & 5469 & 35 & 0.74 & 0.01 & 1.02 & 0.02 & 4.47 & 0.03 & 0.96 & 0.02\\
  HD 43691 & 3.1 & 2.5 & 5920 & 34 & 2.24 & 0.02 & 1.21 & 0.04 & 4.19 & 0.02 & 1.44 & 0.03\\
  HD 44219 & 9.6 & 0.7 & 5749 & 45 & 1.83 & 0.01 & 1.01 & 0.01 & 4.17 & 0.02 & 1.37 & 0.03\\
  HD 45364 & 3.4 & 2.7 & 5540 & 31 & 0.562 & 0.004 & 0.88 & 0.02 & 4.55 & 0.03 & 0.82 & 0.01\\
  CoRoT-19 & 0.7 & 0.7 & 6066 & 23 & 1.239 & 0.003 & 1.09 & 0.01 & 4.46 & 0.01 & 1.01 & 0.01\\
  HD 45350 & 7.0 & 0.9 & 5683 & 39 & 1.43 & 0.01 & 1.06 & 0.02 & 4.27 & 0.03 & 1.24 & 0.02\\
  HD 45652 & 6.0 & 2.9 & 5348 & 38 & 0.62 & 0.01 & 0.94 & 0.03 & 4.48 & 0.03 & 0.92 & 0.02\\
  6 Lyn & 2.8 & 0.2 & 4994 & 15 & 14.9 & 0.1 & 1.46 & 0.02 & 3.17 & 0.03 & 5.2 & 0.1\\
  CoRoT-18 & 0.03 & 0.34 & 5364 & 18 & 0.7 & 0.003 & 0.94 & 0.01 & 4.429 & 0.005 & 0.97 & 0.01\\
  HD 47186 & 5.5 & 0.6 & 5736 & 21 & 1.219 & 0.005 & 1.05 & 0.01 & 4.35 & 0.01 & 1.12 & 0.01\\
  7 CMa & 4.6 & 0.7 & 4790 & 27 & 11.3 & 0.1 & 1.3 & 0.1 & 3.18 & 0.03 & 4.9 & 0.1\\
  CoRoT-12 & 12.5 & 0.4 & 5370 & 7 & 0.753 & 0.001 & 0.894 & 0.004 & 4.38 & 0.02 & 1.0 & 0.02\\
  CoRoT-7 & 6.9 & 2.7 & 5226 & 22 & 0.462 & 0.003 & 0.86 & 0.02 & 4.53 & 0.03 & 0.83 & 0.01\\
  CoRoT-13 & 12.5 & 0.4 & 5605 & 52 & 1.45 & 0.01 & 0.95 & 0.02 & 4.2 & 0.01 & 1.26 & 0.01\\
  HD 49674 & 1.8 & 1.2 & 5702 & 28 & 0.96 & 0.01 & 1.07 & 0.02 & 4.46 & 0.02 & 1.01 & 0.01\\
  HD 50499 & 2.7 & 0.6 & 6102 & 54 & 2.25 & 0.02 & 1.25 & 0.01 & 4.27 & 0.03 & 1.35 & 0.03\\
  CoRoT-14 & 0.03 & 0.13 & 5204 & 32 & 0.67 & 0.01 & 0.98 & 0.01 & 4.42 & 0.01 & 1.01 & 0.02\\
  HD 50554 & 3.3 & 1.4 & 6036 & 52 & 1.37 & 0.01 & 1.06 & 0.03 & 4.4 & 0.04 & 1.07 & 0.03\\
  HD 52265 & 2.6 & 0.6 & 6163 & 41 & 2.08 & 0.01 & 1.21 & 0.02 & 4.31 & 0.03 & 1.27 & 0.03\\
  HAT-P-24 & 2.4 & 0.7 & 6487 & 47 & 2.54 & 0.01 & 1.18 & 0.02 & 4.3 & 0.03 & 1.26 & 0.02\\
  HAT-P-9 & 0.8 & 0.8 & 6358 & 40 & 2.17 & 0.01 & 1.25 & 0.03 & 4.36 & 0.02 & 1.21 & 0.01\\
  HAT-P-33 & 4.2 & 0.6 & 5843 & 110 & 5.4 & 0.7 & 1.3 & 0.1 & 3.9 & 0.1 & 2.1 & 0.3\\
  HD 63454 & 2.7 & 3.3 & 4788 & 20 & 0.242 & 0.003 & 0.79 & 0.02 & 4.62 & 0.02 & 0.72 & 0.01\\
  XO-5 & 12.1 & 0.9 & 5409 & 40 & 0.93 & 0.01 & 0.93 & 0.01 & 4.33 & 0.01 & 1.08 & 0.01\\
  HD 63765 & 7.2 & 3.6 & 5483 & 42 & 0.58 & 0.01 & 0.85 & 0.03 & 4.51 & 0.04 & 0.84 & 0.02\\
  XO-2 & 2.1 & 1.4 & 5412 & 21 & 0.671 & 0.004 & 1.01 & 0.02 & 4.5 & 0.02 & 0.93 & 0.01\\
  HD 65216 & 1.7 & 0.5 & 5718 & 8 & 0.716 & 0.001 & 0.95 & 0.01 & 4.53 & 0.01 & 0.864 & 0.003\\
  HD 66428 & 4.1 & 1.4 & 5773 & 55 & 1.28 & 0.01 & 1.09 & 0.02 & 4.37 & 0.03 & 1.13 & 0.03\\
  HAT-P-30 & 0.8 & 0.2 & 6329 & 11 & 2.069 & 0.002 & 1.24 & 0.01 & 4.37 & 0.01 & 1.199 & 0.005\\
  HD 68988 & 1.0 & 0.4 & 5919 & 11 & 1.297 & 0.002 & 1.16 & 0.01 & 4.42 & 0.01 & 1.08 & 0.01\\
  HD 69830 & 11.2 & 1.7 & 5396 & 15 & 0.596 & 0.002 & 0.84 & 0.02 & 4.47 & 0.01 & 0.88 & 0.01\\
  HD 70642 & 1.9 & 1.1 & 5732 & 23 & 0.917 & 0.004 & 1.04 & 0.02 & 4.47 & 0.02 & 0.97 & 0.01\\
  HD 72659 & 7.0 & 0.7 & 5956 & 43 & 2.16 & 0.01 & 1.07 & 0.02 & 4.19 & 0.02 & 1.38 & 0.02\\
  HD 73256 & 2.5 & 2.3 & 5532 & 36 & 0.74 & 0.01 & 1.01 & 0.03 & 4.49 & 0.03 & 0.94 & 0.02\\
  HD 73526 & 7.9 & 0.3 & 5675 & 33 & 2.21 & 0.01 & 1.09 & 0.01 & 4.1 & 0.03 & 1.54 & 0.04\\
  HD 73534 & 5.8 & 0.9 & 5006 & 44 & 3.41 & 0.04 & 1.22 & 0.05 & 3.71 & 0.03 & 2.5 & 0.1\\
  HAT-P-13 & 6.2 & 0.8 & 5731 & 35 & 1.95 & 0.01 & 1.14 & 0.03 & 4.19 & 0.02 & 1.42 & 0.02\\
  4 UMa & 12.3 & 0.7 & 4554 & 59 & 37.0 & 1.1 & 0.94 & 0.03 & 2.45 & 0.01 & 9.4 & 0.03\\
  HD 74156 & 4.3 & 0.6 & 6074 & 48 & 3.08 & 0.01 & 1.24 & 0.03 & 4.12 & 0.02 & 1.59 & 0.03\\
  WASP-36 & 0.7 & 0.2 & 6333 & 7 & 1.4292 & 5.0E-4 & 1.061 & 0.004 & 4.461 & 0.004 & 0.995 & 0.002\\
  HD 75898 & 4.0 & 0.4 & 5998 & 38 & 2.24 & 0.01 & 1.21 & 0.01 & 4.23 & 0.03 & 1.39 & 0.03\\
  HD 76700 & 6.9 & 0.8 & 5694 & 44 & 1.69 & 0.01 & 1.1 & 0.02 & 4.22 & 0.03 & 1.34 & 0.03\\
  WASP-13 & 12.59 & 0.01 & 5625 & 42 & 1.44 & 0.02 & 0.94 & 0.02 & 4.2072 & 1.0E-4 & 1.241 & 1.0E-4\\
  HD 81040 & 5.5 & 1.8 & 5707 & 31 & 0.78 & 0.005 & 0.92 & 0.02 & 4.48 & 0.03 & 0.91 & 0.01\\
  HD 81688 & 6.5 & 3.2 & 4847 & 55 & 54.5 & 1.0 & 1.1 & 0.2 & 2.4 & 0.1 & 10.5 & 0.3\\
  HD 82943 & 1.3 & 0.5 & 6022 & 21 & 1.503 & 0.004 & 1.18 & 0.01 & 4.4 & 0.01 & 1.13 & 0.01\\
  HD 82886 & 3.4 & 0.6 & 5105 & 45 & 11.9 & 0.1 & 1.3 & 0.1 & 3.26 & 0.04 & 4.4 & 0.1\\
  HD 83443 & 3.3 & 1.6 & 5497 & 28 & 0.77 & 0.01 & 1.02 & 0.02 & 4.47 & 0.02 & 0.97 & 0.01\\
  HD 85390 & 5.6 & 3.7 & 5186 & 26 & 0.389 & 0.004 & 0.82 & 0.02 & 4.57 & 0.03 & 0.77 & 0.01\\
  HD 85512 & 8.2 & 3.3 & 4555 & 6 & 0.1339 & 5.0E-4 & 0.62 & 0.01 & 4.68 & 0.01 & 0.589 & 0.003\\
  HD 86081 & 5.2 & 0.7 & 5905 & 40 & 2.43 & 0.01 & 1.19 & 0.03 & 4.16 & 0.03 & 1.5 & 0.03\\
  HD 86264 & 1.4 & 0.1 & 6589 & 15 & 4.011 & 0.005 & 1.416 & 0.003 & 4.21 & 0.01 & 1.54 & 0.01\\
  BD -08 2823 & 8.7 & 3.5 & 4721 & 12 & 0.211 & 0.001 & 0.72 & 0.01 & 4.61 & 0.02 & 0.69 & 0.01\\
  HD 87883 & 9.1 & 3.5 & 4956 & 24 & 0.325 & 0.004 & 0.8 & 0.02 & 4.56 & 0.02 & 0.77 & 0.01\\
  HD 88133 & 5.0 & 0.1 & 5466 & 28 & 3.7 & 0.1 & 1.28 & 0.01 & 3.87 & 0.02 & 2.1 & 0.1\\
  HD 89307 & 6.3 & 0.3 & 5960 & 12 & 1.377 & 0.002 & 0.99 & 0.01 & 4.34 & 0.02 & 1.1 & 0.01\\
  WASP-43 & 0.2 & 0.6 & 4371 & 8 & 0.104 & 0.001 & 0.63 & 0.01 & 4.73 & 0.01 & 0.563 & 0.004\\
  HAT-P-22 & 12.2 & 0.9 & 5358 & 37 & 0.79 & 0.01 & 0.91 & 0.01 & 4.37 & 0.01 & 1.01 & 0.01\\
  24 Sex & 2.3 & 0.3 & 5048 & 44 & 13.6 & 0.1 & 1.6 & 0.1 & 3.27 & 0.03 & 4.8 & 0.1\\
  HD 90156 & 5.9 & 1.5 & 5719 & 24 & 0.746 & 0.003 & 0.89 & 0.02 & 4.49 & 0.02 & 0.88 & 0.01\\
  HD 92788 & 2.3 & 1.3 & 5838 & 47 & 1.25 & 0.01 & 1.12 & 0.02 & 4.4 & 0.03 & 1.1 & 0.02\\
  HD 93083 & 5.4 & 4.7 & 5035 & 43 & 0.38 & 0.01 & 0.85 & 0.03 & 4.54 & 0.03 & 0.81 & 0.02\\
  BD -10 3166 & 5.2 & 3.4 & 5257 & 40 & 0.56 & 0.01 & 0.94 & 0.03 & 4.5 & 0.03 & 0.9 & 0.02\\
  HD 95089 & 2.3 & 0.2 & 5007 & 37 & 13.2 & 0.1 & 1.61 & 0.05 & 3.27 & 0.03 & 4.8 & 0.1\\
  47 UMa & 5.1 & 1.0 & 5951 & 48 & 1.61 & 0.01 & 1.08 & 0.02 & 4.31 & 0.04 & 1.19 & 0.03\\
  WASP-34 & 12.2 & 0.8 & 5567 & 45 & 0.98 & 0.01 & 0.91 & 0.02 & 4.34 & 0.01 & 1.05 & 0.01\\
  HD 96063 & 3.6 & 0.7 & 5103 & 47 & 11.0 & 0.1 & 1.3 & 0.1 & 3.29 & 0.04 & 4.3 & 0.1\\
  HD 96167 & 5.1 & 0.3 & 5665 & 20 & 3.8 & 0.1 & 1.26 & 0.03 & 3.93 & 0.04 & 2.0 & 0.1\\
  HD 96127 & 5.4 & 2.4 & 3955 & 35 & 503.4 & 20.7 & 1.2 & 0.2 & 1.1 & 0.1 & 48.1 & 1.8\\
  HD 97658 & 3.8 & 2.6 & 5190 & 17 & 0.34 & 0.002 & 0.78 & 0.02 & 4.61 & 0.02 & 0.72 & 0.01\\
  HD 98219 & 4.0 & 0.8 & 4960 & 46 & 9.6 & 0.2 & 1.3 & 0.1 & 3.31 & 0.05 & 4.2 & 0.1\\
  HD 99109 & 4.1 & 2.7 & 5268 & 30 & 0.54 & 0.01 & 0.94 & 0.02 & 4.51 & 0.03 & 0.88 & 0.02\\
  HD 99706 & 2.8 & 0.2 & 4904 & 16 & 13.1 & 0.1 & 1.53 & 0.03 & 3.21 & 0.02 & 5.03 & 0.05\\
  HD 100655 & 0.9 & 0.2 & 4918 & 8 & 40.8 & 0.3 & 2.2 & 0.1 & 2.89 & 0.02 & 8.8 & 0.1\\
  HD 100777 & 5.3 & 1.3 & 5590 & 29 & 0.92 & 0.01 & 1.01 & 0.02 & 4.41 & 0.02 & 1.02 & 0.01\\
  HD 101930 & 5.4 & 4.4 & 5147 & 39 & 0.43 & 0.01 & 0.87 & 0.03 & 4.54 & 0.03 & 0.83 & 0.02\\
  HIP 57274 & 8.4 & 3.7 & 4660 & 11 & 0.211 & 0.001 & 0.74 & 0.02 & 4.6 & 0.02 & 0.71 & 0.01\\
  HD 102117 & 5.5 & 0.9 & 5740 & 41 & 1.38 & 0.01 & 1.08 & 0.02 & 4.32 & 0.02 & 1.19 & 0.02\\
  HD 102195 & 5.5 & 3.5 & 5281 & 32 & 0.489 & 0.005 & 0.88 & 0.03 & 4.53 & 0.03 & 0.84 & 0.01\\
  HD 102272 & 9.2 & 1.9 & 4778 & 37 & 21.9 & 0.3 & 1.0 & 0.1 & 2.77 & 0.05 & 6.8 & 0.2\\
  HD 102329 & 2.0 & 0.3 & 4867 & 42 & 16.1 & 0.2 & 1.8 & 0.1 & 3.16 & 0.04 & 5.7 & 0.1\\
  HD 102956 & 2.4 & 0.2 & 5016 & 50 & 9.9 & 0.1 & 1.62 & 0.04 & 3.4 & 0.04 & 4.2 & 0.1\\
  HD 103197 & 2.5 & 2.0 & 5236 & 18 & 0.467 & 0.003 & 0.92 & 0.02 & 4.55 & 0.02 & 0.83 & 0.01\\
  HD 104067 & 10.1 & 2.9 & 4974 & 14 & 0.311 & 0.002 & 0.77 & 0.01 & 4.56 & 0.02 & 0.75 & 0.01\\
  HD 106252 & 7.2 & 1.0 & 5870 & 36 & 1.31 & 0.01 & 0.99 & 0.02 & 4.34 & 0.03 & 1.11 & 0.02\\
  HD 106270 & 4.0 & 0.1 & 5498 & 32 & 5.2 & 0.1 & 1.35 & 0.02 & 3.76 & 0.03 & 2.5 & 0.1\\
  HD 107148 & 4.3 & 1.2 & 5795 & 55 & 1.38 & 0.01 & 1.11 & 0.02 & 4.34 & 0.03 & 1.17 & 0.03\\
  HD 108147 & 1.7 & 0.7 & 6229 & 39 & 1.9 & 0.01 & 1.19 & 0.02 & 4.36 & 0.02 & 1.18 & 0.02\\
  HD 108863 & 2.2 & 0.2 & 4940 & 31 & 13.9 & 0.2 & 1.68 & 0.04 & 3.24 & 0.03 & 5.1 & 0.1\\
  HD 108874 & 7.5 & 0.8 & 5630 & 24 & 1.066 & 0.005 & 1.0 & 0.01 & 4.36 & 0.03 & 1.09 & 0.02\\
  HD 109246 & 2.5 & 0.7 & 5890 & 21 & 1.155 & 0.004 & 1.07 & 0.01 & 4.43 & 0.02 & 1.03 & 0.01\\
  HAT-P-36 & 11.3 & 1.2 & 5467 & 44 & 1.12 & 0.01 & 0.96 & 0.01 & 4.27 & 0.03 & 1.18 & 0.03\\
  WASP-41 & 4.0 & 2.2 & 5564 & 28 & 0.642 & 0.004 & 0.91 & 0.02 & 4.52 & 0.03 & 0.86 & 0.01\\
  HD 111232 & 11.7 & 1.4 & 5648 & 30 & 0.7 & 0.003 & 0.8 & 0.02 & 4.45 & 0.02 & 0.88 & 0.01\\
  HD 114386 & 6.6 & 3.8 & 4913 & 18 & 0.286 & 0.003 & 0.78 & 0.02 & 4.59 & 0.02 & 0.74 & 0.01\\
  HD 114783 & 9.8 & 3.0 & 5074 & 23 & 0.397 & 0.003 & 0.83 & 0.02 & 4.53 & 0.02 & 0.81 & 0.01\\
  HD 116029 & 3.5 & 0.5 & 4906 & 40 & 9.7 & 0.1 & 1.4 & 0.1 & 3.34 & 0.03 & 4.3 & 0.1\\
  70 Vir & 8.0 & 0.1 & 5618 & 66 & 2.9 & 0.1 & 1.08 & 0.01 & 3.92 & 0.02 & 1.88 & 0.05\\
  HD 117207 & 4.5 & 1.7 & 5732 & 53 & 1.16 & 0.01 & 1.06 & 0.03 & 4.38 & 0.03 & 1.09 & 0.03\\
  HD 117618 & 5.0 & 0.8 & 5995 & 39 & 1.63 & 0.01 & 1.08 & 0.02 & 4.32 & 0.03 & 1.19 & 0.02\\
  HAT-P-3 & 0.9 & 0.3 & 5221 & 5 & 0.455 & 0.001 & 0.931 & 0.004 & 4.567 & 0.003 & 0.824 & 0.002\\
  HD 121504 & 1.7 & 0.9 & 6088 & 41 & 1.61 & 0.01 & 1.17 & 0.02 & 4.38 & 0.02 & 1.14 & 0.02\\
  HAT-P-26 & 8.0 & 1.8 & 5093 & 11 & 0.359 & 0.001 & 0.8 & 0.01 & 4.56 & 0.01 & 0.771 & 0.005\\
  HD 125595 & 10.0 & 3.0 & 4708 & 16 & 0.223 & 0.002 & 0.74 & 0.02 & 4.6 & 0.01 & 0.704 & 0.005\\
  WASP-39 & 12.0 & 0.9 & 5470 & 11 & 0.68 & 0.01 & 0.85 & 0.01 & 4.43 & 0.01 & 0.92 & 0.02\\
  WASP-14 & 1.5 & 0.6 & 6499 & 45 & 2.61 & 0.01 & 1.25 & 0.02 & 4.31 & 0.02 & 1.28 & 0.02\\
  HD 128311 & 1.7 & 2.5 & 4917 & 17 & 0.296 & 0.003 & 0.84 & 0.01 & 4.6 & 0.02 & 0.75 & 0.01\\
  HD 130322 & 4.6 & 3.3 & 5425 & 35 & 0.557 & 0.005 & 0.9 & 0.03 & 4.53 & 0.02 & 0.85 & 0.01\\
  WASP-37 & 6.2 & 2.7 & 5879 & 51 & 0.84 & 0.01 & 0.87 & 0.03 & 4.48 & 0.04 & 0.89 & 0.02\\
  HD 131496 & 4.5 & 0.4 & 4827 & 19 & 8.6 & 0.1 & 1.34 & 0.03 & 3.31 & 0.02 & 4.2 & 0.05\\
  HD 131664 & 2.6 & 1.0 & 5921 & 42 & 1.47 & 0.01 & 1.15 & 0.02 & 4.37 & 0.03 & 1.15 & 0.02\\
  HD 134987 & 4.7 & 0.7 & 5822 & 39 & 1.49 & 0.01 & 1.11 & 0.01 & 4.32 & 0.03 & 1.2 & 0.02\\
  11 UMi & 3.9 & 0.9 & 4140 & 16 & 243.4 & 3.6 & 1.4 & 0.1 & 1.61 & 0.04 & 30.4 & 0.5\\
  HD 136118 & 5.3 & 0.6 & 6135 & 37 & 3.03 & 0.01 & 1.15 & 0.03 & 4.12 & 0.03 & 1.54 & 0.03\\
  HD 136418 & 4.7 & 0.6 & 4987 & 29 & 6.9 & 0.1 & 1.26 & 0.05 & 3.44 & 0.03 & 3.5 & 0.1\\
  HAT-P-4 & 6.1 & 0.7 & 5805 & 57 & 2.74 & 0.01 & 1.17 & 0.04 & 4.06 & 0.03 & 1.65 & 0.04\\
  HD 137510 & 3.1 & 0.2 & 6032 & 44 & 4.33 & 0.01 & 1.41 & 0.01 & 4.02 & 0.02 & 1.91 & 0.03\\
  HD 139357 & 7.0 & 2.0 & 4457 & 41 & 73.5 & 1.7 & 1.1 & 0.1 & 2.2 & 0.1 & 14.4 & 0.4\\
  HD 137388 & 7.4 & 3.9 & 5182 & 35 & 0.47 & 0.01 & 0.88 & 0.03 & 4.51 & 0.03 & 0.85 & 0.02\\
  HD 330075 & 0.1 & 0.5 & 4977 & 45 & 0.46 & 0.01 & 0.91 & 0.02 & 4.47 & 0.02 & 0.91 & 0.03\\
  HD 141937 & 1.2 & 0.7 & 5891 & 18 & 1.108 & 0.003 & 1.09 & 0.01 & 4.46 & 0.01 & 1.01 & 0.01\\
  HD 142245 & 3.3 & 0.4 & 4858 & 34 & 11.4 & 0.1 & 1.5 & 0.1 & 3.24 & 0.03 & 4.8 & 0.1\\
  HD 142415 & 1.5 & 0.9 & 5918 & 24 & 1.153 & 0.005 & 1.09 & 0.02 & 4.45 & 0.02 & 1.02 & 0.01\\
  rho CrB & 9.1 & 1.0 & 5911 & 54 & 1.76 & 0.01 & 0.97 & 0.02 & 4.21 & 0.05 & 1.27 & 0.04\\
  XO-1 & 2.2 & 1.4 & 5742 & 27 & 0.849 & 0.004 & 1.0 & 0.02 & 4.49 & 0.02 & 0.93 & 0.01\\
  HD 145457 & 2.6 & 0.4 & 4785 & 37 & 40.5 & 0.5 & 1.5 & 0.1 & 2.67 & 0.04 & 9.3 & 0.2\\
  14 Her & 3.6 & 2.0 & 5349 & 29 & 0.64 & 0.01 & 0.98 & 0.02 & 4.48 & 0.02 & 0.93 & 0.01\\
  HD 145377 & 2.6 & 1.3 & 5987 & 50 & 1.42 & 0.01 & 1.12 & 0.03 & 4.39 & 0.03 & 1.11 & 0.02\\
  WASP-38 & 0.4 & 0.5 & 6321 & 22 & 1.624 & 0.003 & 1.14 & 0.02 & 4.43 & 0.01 & 1.06 & 0.01\\
  HAT-P-2 & 0.8 & 0.2 & 6671 & 23 & 3.42 & 0.01 & 1.37 & 0.01 & 4.28 & 0.01 & 1.39 & 0.01\\
  HD 147018 & 7.4 & 1.8 & 5528 & 31 & 0.78 & 0.01 & 0.94 & 0.02 & 4.43 & 0.04 & 0.97 & 0.02\\
  HD 148156 & 1.0 & 0.4 & 6150 & 20 & 1.822 & 0.004 & 1.23 & 0.01 & 4.37 & 0.01 & 1.19 & 0.01\\
  HD 148427 & 3.55 & 0.04 & 5029 & 15 & 6.05 & 0.03 & 1.43 & 0.01 & 3.56 & 0.01 & 3.25 & 0.03\\
  HD 149026 & 2.9 & 0.3 & 6110 & 50 & 2.78 & 0.02 & 1.301 & 0.005 & 4.2 & 0.02 & 1.49 & 0.03\\
  HD 150706 & 0.3 & 0.3 & 5958 & 8 & 1.043 & 0.001 & 1.07 & 0.01 & 4.49 & 0.01 & 0.96 & 0.003\\
  HD 149143 & 7.1 & 0.3 & 5745 & 31 & 2.28 & 0.01 & 1.12 & 0.01 & 4.11 & 0.03 & 1.53 & 0.04\\
  HD 152581 & 8.6 & 2.1 & 4995 & 44 & 13.5 & 0.2 & 1.0 & 0.1 & 3.0 & 0.1 & 4.9 & 0.1\\
  HD 154345 & 3.1 & 0.4 & 5558 & 5 & 0.609 & 0.001 & 0.907 & 0.004 & 4.54 & 0.01 & 0.843 & 0.002\\
  HD 153950 & 4.5 & 0.6 & 6111 & 53 & 2.24 & 0.01 & 1.15 & 0.01 & 4.24 & 0.04 & 1.34 & 0.03\\
  HD 155358 & 0.3 & 2.0 & 5933 & 49 & 2.1 & 0.01 & 1.11 & 0.05 & 4.2 & 0.02 & 1.37 & 0.03\\
  HD 154672 & 7.0 & 0.6 & 5754 & 37 & 1.85 & 0.01 & 1.1 & 0.01 & 4.2 & 0.03 & 1.37 & 0.03\\
  HD 154857 & 5.6 & 0.2 & 5729 & 36 & 4.5 & 0.1 & 1.14 & 0.01 & 3.81 & 0.01 & 2.17 & 0.04\\
  HD 156279 & 7.4 & 1.9 & 5453 & 30 & 0.71 & 0.01 & 0.93 & 0.02 & 4.45 & 0.03 & 0.95 & 0.01\\
  HD 156668 & 8.8 & 3.5 & 4859 & 15 & 0.278 & 0.002 & 0.78 & 0.02 & 4.58 & 0.02 & 0.75 & 0.01\\
  HD 156411 & 4.4 & 0.2 & 5921 & 64 & 5.4 & 0.1 & 1.23 & 0.02 & 3.84 & 0.02 & 2.2 & 0.1\\
  HAT-P-14 & 0.7 & 0.4 & 6684 & 43 & 3.41 & 0.01 & 1.37 & 0.02 & 4.29 & 0.02 & 1.38 & 0.02\\
  HD 158038 & 3.7 & 0.5 & 4837 & 38 & 10.2 & 0.1 & 1.4 & 0.1 & 3.27 & 0.04 & 4.6 & 0.1\\
  HD 159868 & 6.1 & 0.4 & 5593 & 54 & 3.8 & 0.1 & 1.16 & 0.02 & 3.87 & 0.02 & 2.1 & 0.1\\
  mu Ara & 5.7 & 0.6 & 5815 & 40 & 1.81 & 0.01 & 1.13 & 0.02 & 4.24 & 0.03 & 1.33 & 0.02\\
  HD 162020 & 1.7 & 1.5 & 4756 & 7 & 0.23 & 0.001 & 0.79 & 0.01 & 4.63 & 0.01 & 0.708 & 0.004\\
  TrES-3 & 1.8 & 1.5 & 5613 & 18 & 0.581 & 0.003 & 0.89 & 0.01 & 4.57 & 0.01 & 0.81 & 0.01\\
  TrES-4 & 2.8 & 0.1 & 6308 & 24 & 4.39 & 0.01 & 1.374 & 0.001 & 4.08 & 0.02 & 1.76 & 0.04\\
  HD 163607 & 8.91 & 0.01 & 5541 & 2 & 2.316 & 0.001 & 1.0719 & 2.0E-4 & 4.024 & 0.002 & 1.65 & 0.01\\
  HD 164509 & 1.5 & 0.2 & 5865 & 7 & 1.15 & 0.001 & 1.103 & 0.004 & 4.44 & 0.01 & 1.041 & 0.003\\
  HD 164922 & 5.2 & 2.3 & 5467 & 34 & 0.69 & 0.01 & 0.95 & 0.02 & 4.48 & 0.03 & 0.93 & 0.02\\
  HAT-P-31 & 0.6 & 0.1 & 6380 & 5 & 2.2391 & 4.0E-4 & 1.269 & 0.003 & 4.357 & 0.002 & 1.227 & 0.002\\
  HAT-P-5 & 11.5 & 1.0 & 5005 & 20 & 2.0 & 0.1 & 1.02 & 0.03 & 3.88 & 0.04 & 1.9 & 0.1\\
  HD 168443 & 10.8 & 0.6 & 5585 & 83 & 2.082 & 0.003 & 1.0 & 0.01 & 4.05 & 0.04 & 1.5 & 0.1\\
  HD 168746 & 11.2 & 1.4 & 5661 & 37 & 1.05 & 0.01 & 0.91 & 0.02 & 4.33 & 0.02 & 1.07 & 0.02\\
  42 Dra & 9.0 & 2.1 & 4384 & 31 & 146.6 & 2.7 & 1.0 & 0.1 & 1.8 & 0.1 & 20.9 & 0.6\\
  HD 169830 & 2.82 & 0.03 & 6300 & 23 & 4.703 & 0.005 & 1.39 & 0.01 & 4.06 & 0.01 & 1.81 & 0.02\\
  HD 170469 & 8.6 & 0.5 & 4990 & 28 & 2.5 & 0.1 & 1.11 & 0.02 & 3.84 & 0.03 & 2.1 & 0.1\\
  HD 171028 & 6.1 & 0.5 & 5787 & 81 & 4.9 & 0.2 & 1.06 & 0.03 & 3.76 & 0.03 & 2.3 & 0.1\\
  CoRoT-16 & 12.1 & 1.0 & 5021 & 16 & 0.395 & 0.002 & 0.82 & 0.01 & 4.51 & 0.01 & 0.826 & 0.003\\
  HD 171238 & 2.5 & 1.6 & 5596 & 26 & 0.762 & 0.004 & 1.0 & 0.02 & 4.5 & 0.02 & 0.93 & 0.01\\
  CoRoT-23 & 0.0033 & 5.0E-4 & 3704 & 32 & 0.201 & 0.003 & 0.4 & 0.02 & 3.96 & 0.04 & 1.09 & 0.03\\
  CoRoT-11 & 1.5 & 4.0 & 4994 & 39 & 2.42 & 0.04 & 1.5 & 0.2 & 4.0 & 0.1 & 2.07 & 0.05\\
  CoRoT-9 & 12.1 & 1.0 & 5380 & 40 & 0.64 & 0.01 & 0.86 & 0.02 & 4.44 & 0.01 & 0.909 & 0.005\\
  HD 173416 & 1.5 & 0.6 & 4777 & 21 & 80.5 & 0.5 & 1.8 & 0.1 & 2.46 & 0.04 & 13.1 & 0.2\\
  HD 175541 & 2.51 & 0.03 & 5162 & 17 & 9.55 & 0.02 & 1.54 & 0.02 & 3.44 & 0.01 & 3.85 & 0.04\\
  TrES-1 & 1.5 & 1.1 & 5281 & 10 & 0.434 & 0.002 & 0.88 & 0.01 & 4.58 & 0.01 & 0.789 & 0.005\\
  HD 177830 & 5.3 & 0.9 & 4789 & 48 & 5.2 & 0.1 & 1.3 & 0.1 & 3.51 & 0.05 & 3.3 & 0.1\\
  TrES-2 & 5.1 & 0.8 & 5891 & 19 & 1.053 & 0.003 & 0.96 & 0.01 & 4.43 & 0.01 & 0.99 & 0.01\\
  Kepler-21 & 3.55 & 0.03 & 6203 & 5 & 5.281 & 0.001 & 1.311 & 0.003 & 3.95 & 0.05 & 1.99 & 0.04\\
  Kepler-20 & 12.5 & 0.3 & 5432 & 42 & 0.74 & 0.01 & 0.88 & 0.02 & 4.403 & 0.004 & 0.958 & 0.002\\
  HD 179079 & 7.3 & 0.4 & 5678 & 72 & 2.43 & 0.01 & 1.12 & 0.01 & 4.07 & 0.03 & 1.6 & 0.1\\
  HD 180314 & 0.9 & 0.6 & 4977 & 31 & 39.8 & 0.7 & 2.3 & 0.1 & 2.94 & 0.04 & 8.5 & 0.2\\
  HD 179949 & 1.9 & 0.3 & 6215 & 15 & 1.954 & 0.003 & 1.2 & 0.01 & 4.35 & 0.01 & 1.21 & 0.01\\
  HD 180902 & 3.0 & 0.2 & 5026 & 27 & 9.3 & 0.1 & 1.49 & 0.03 & 3.39 & 0.02 & 4.0 & 0.1\\
  HD 181342 & 3.6 & 0.5 & 4992 & 49 & 8.4 & 0.1 & 1.4 & 0.1 & 3.4 & 0.04 & 3.9 & 0.1\\
  HD 181720 & 12.6 & 0.2 & 5888 & 108 & 1.91 & 0.02 & 0.86 & 0.02 & 4.134 & 0.002 & 1.29 & 0.003\\
  HD 181433 & 8.6 & 3.6 & 4912 & 25 & 0.34 & 0.005 & 0.83 & 0.02 & 4.54 & 0.02 & 0.8 & 0.01\\
  CoRoT-3 & 1.2 & 2.8 & 5066 & 51 & 3.3 & 0.1 & 1.6 & 0.2 & 3.9 & 0.1 & 2.4 & 0.1\\
  HD 183263 & 4.8 & 0.8 & 5874 & 59 & 1.81 & 0.02 & 1.15 & 0.02 & 4.26 & 0.03 & 1.3 & 0.04\\
  HAT-P-7 & 1.5 & 0.2 & 6659 & 49 & 5.24 & 0.01 & 1.49 & 0.01 & 4.13 & 0.02 & 1.72 & 0.03\\
  HD 231701 & 3.5 & 0.2 & 5873 & 63 & 7.3 & 0.1 & 1.41 & 0.03 & 3.75 & 0.03 & 2.6 & 0.1\\
  HD 187123 & 5.6 & 0.9 & 5855 & 40 & 1.45 & 0.01 & 1.06 & 0.02 & 4.32 & 0.02 & 1.17 & 0.02\\
  Kepler-40 & 0.04 & 0.02 & 7744 & 8 & 7.25 & 0.01 & 1.656 & 0.002 & 4.298 & 0.002 & 1.499 & 0.004\\
  HD 187085 & 3.6 & 0.3 & 6117 & 27 & 2.06 & 0.01 & 1.17 & 0.01 & 4.28 & 0.02 & 1.28 & 0.02\\
  HAT-P-11 & 1.6 & 1.5 & 4747 & 9 & 0.242 & 0.001 & 0.82 & 0.01 & 4.62 & 0.01 & 0.729 & 0.004\\
  Kepler-17 & 0.6 & 2.5 & 5407 & 53 & 1.78 & 0.02 & 1.4 & 0.1 & 4.2 & 0.03 & 1.53 & 0.04\\
  xi Aql & 2.4 & 0.7 & 4796 & 17 & 45.1 & 0.3 & 1.5 & 0.1 & 2.64 & 0.03 & 9.7 & 0.1\\
  HD 189733 & 6.2 & 3.4 & 5038 & 21 & 0.327 & 0.003 & 0.8 & 0.02 & 4.58 & 0.02 & 0.75 & 0.01\\
  HD 190228 & 5.0 & 0.5 & 5352 & 30 & 4.4 & 0.2 & 1.18 & 0.04 & 3.73 & 0.02 & 2.4 & 0.1\\
  HD 190647 & 8.3 & 0.5 & 5656 & 60 & 2.19 & 0.02 & 1.08 & 0.01 & 4.09 & 0.03 & 1.54 & 0.04\\
  HAT-P-34 & 0.8 & 0.3 & 6778 & 32 & 4.46 & 0.01 & 1.47 & 0.03 & 4.23 & 0.01 & 1.54 & 0.02\\
  Qatar-1 & 8.9 & 3.7 & 4757 & 16 & 0.256 & 0.002 & 0.78 & 0.02 & 4.58 & 0.02 & 0.74 & 0.01\\
  HD 192263 & 1.8 & 0.8 & 4965 & 7 & 0.299 & 0.001 & 0.821 & 0.005 & 4.61 & 0.01 & 0.74 & 0.003\\
  GJ 785 & 1.0 & 0.6 & 5179 & 11 & 0.396 & 0.003 & 0.88 & 0.005 & 4.589 & 0.004 & 0.782 & 0.002\\
  HD 192310 & 10.0 & 2.5 & 5136 & 17 & 0.398 & 0.003 & 0.81 & 0.02 & 4.53 & 0.02 & 0.8 & 0.01\\
  HD 192699 & 2.5 & 0.1 & 5136 & 15 & 11.12 & 0.03 & 1.5 & 0.01 & 3.36 & 0.01 & 4.22 & 0.03\\
  TrES-5 & 9.2 & 1.8 & 5087 & 14 & 0.42 & 0.002 & 0.85 & 0.01 & 4.51 & 0.01 & 0.84 & 0.01\\
  HAT-P-23 & 12.6 & 0.2 & 4830 & 5 & 2.829 & 0.005 & 0.987 & 0.005 & 3.663 & 0.005 & 2.41 & 0.01\\
  WASP-2 & 0.015 & 0.002 & 4747 & 46 & 0.56 & 0.01 & 1.05 & 0.02 & 4.36 & 0.02 & 1.11 & 0.04\\
  HD 197037 & 0.3 & 0.3 & 6357 & 15 & 1.561 & 0.002 & 1.11 & 0.01 & 4.45 & 0.004 & 1.034 & 0.004\\
  WASP-7 & 0.2 & 0.2 & 6562 & 13 & 2.434 & 0.002 & 1.27 & 0.01 & 4.37 & 0.01 & 1.21 & 0.01\\
  18 Del & 1.1 & 0.1 & 5005 & 35 & 35.7 & 0.3 & 2.0 & 0.1 & 2.94 & 0.03 & 8.0 & 0.2\\
  HD 200964 & 2.7 & 0.3 & 5105 & 42 & 11.3 & 0.1 & 1.47 & 0.05 & 3.33 & 0.03 & 4.3 & 0.1\\
  BD +14 4559 & 9.0 & 3.4 & 4980 & 19 & 0.339 & 0.003 & 0.81 & 0.02 & 4.55 & 0.02 & 0.78 & 0.01\\
  HD 202206 & 1.1 & 0.6 & 5745 & 17 & 1.024 & 0.003 & 1.1 & 0.01 & 4.45 & 0.01 & 1.02 & 0.01\\
  HD 204313 & 5.5 & 0.8 & 5761 & 26 & 1.22 & 0.01 & 1.04 & 0.01 & 4.36 & 0.02 & 1.11 & 0.01\\
  HD 204941 & 5.2 & 3.9 & 5079 & 23 & 0.304 & 0.002 & 0.76 & 0.02 & 4.61 & 0.02 & 0.72 & 0.01\\
  HD 205739 & 2.8 & 0.2 & 6276 & 41 & 3.52 & 0.01 & 1.329 & 0.003 & 4.15 & 0.02 & 1.59 & 0.04\\
  HAT-P-17 & 6.9 & 1.9 & 5273 & 11 & 0.465 & 0.002 & 0.85 & 0.02 & 4.54 & 0.02 & 0.82 & 0.01\\
  HD 206610 & 2.1 & 0.3 & 4907 & 36 & 18.1 & 0.2 & 1.7 & 0.1 & 3.12 & 0.03 & 5.9 & 0.1\\
  HD 208487 & 3.1 & 0.6 & 6119 & 32 & 1.78 & 0.01 & 1.14 & 0.01 & 4.34 & 0.03 & 1.19 & 0.02\\
  HD 209458 & 4.0 & 1.2 & 6084 & 63 & 1.77 & 0.01 & 1.11 & 0.02 & 4.3 & 0.1 & 1.2 & 0.04\\
  HD 210277 & 7.9 & 2.0 & 5566 & 50 & 0.99 & 0.01 & 0.98 & 0.02 & 4.37 & 0.04 & 1.07 & 0.03\\
  HD 210702 & 2.1 & 0.1 & 5027 & 32 & 12.8 & 0.1 & 1.69 & 0.03 & 3.31 & 0.02 & 4.7 & 0.1\\
  HD 212771 & 3.0 & 0.4 & 5064 & 45 & 13.5 & 0.1 & 1.4 & 0.1 & 3.22 & 0.04 & 4.8 & 0.1\\
  HD 215497 & 10.1 & 2.8 & 5082 & 24 & 0.43 & 0.004 & 0.85 & 0.02 & 4.5 & 0.02 & 0.85 & 0.01\\
  HAT-P-8 & 0.3 & 0.4 & 6697 & 38 & 3.05 & 0.01 & 1.33 & 0.02 & 4.33 & 0.01 & 1.3 & 0.02\\
  tau Gru & 4.2 & 0.6 & 5996 & 56 & 3.39 & 0.02 & 1.28 & 0.04 & 4.07 & 0.03 & 1.71 & 0.04\\
  HD 216437 & 5.2 & 0.6 & 5909 & 31 & 2.28 & 0.01 & 1.17 & 0.03 & 4.18 & 0.04 & 1.44 & 0.03\\
  HD 216770 & 4.5 & 2.3 & 5413 & 33 & 0.65 & 0.01 & 0.96 & 0.02 & 4.49 & 0.03 & 0.92 & 0.02\\
  51 Peg & 3.3 & 1.2 & 5882 & 48 & 1.35 & 0.01 & 1.1 & 0.02 & 4.38 & 0.03 & 1.12 & 0.03\\
  HAT-P-1 & 1.9 & 0.6 & 6029 & 24 & 1.473 & 0.005 & 1.14 & 0.02 & 4.39 & 0.01 & 1.12 & 0.01\\
  HD 217107 & 3.1 & 1.5 & 5717 & 50 & 1.14 & 0.01 & 1.1 & 0.02 & 4.4 & 0.03 & 1.09 & 0.02\\
  HD 217786 & 6.5 & 0.8 & 6027 & 42 & 1.73 & 0.01 & 1.02 & 0.01 & 4.28 & 0.03 & 1.21 & 0.02\\
  WASP-21 & 5.5 & 2.0 & 6074 & 58 & 1.17 & 0.01 & 0.93 & 0.03 & 4.42 & 0.04 & 0.98 & 0.02\\
  HD 240210 & 10.9 & 1.8 & 4833 & 28 & 11.9 & 0.1 & 1.0 & 0.1 & 3.05 & 0.04 & 4.9 & 0.1\\
  HD 219828 & 5.0 & 0.7 & 5938 & 42 & 2.64 & 0.01 & 1.2 & 0.04 & 4.14 & 0.04 & 1.54 & 0.04\\
  HD 220773 & 6.3 & 0.1 & 5852 & 14 & 3.32 & 0.01 & 1.156 & 0.002 & 4.0 & 0.01 & 1.78 & 0.01\\
  14 And & 3.2 & 2.1 & 4783 & 39 & 56.1 & 0.8 & 1.4 & 0.2 & 2.5 & 0.1 & 10.9 & 0.3\\
  HD 221287 & 0.8 & 0.5 & 6307 & 22 & 1.859 & 0.005 & 1.19 & 0.01 & 4.39 & 0.01 & 1.14 & 0.01\\
  WASP-4 & 0.9 & 0.4 & 5758 & 7 & 0.8 & 0.002 & 1.0 & 0.01 & 4.52 & 0.01 & 0.901 & 0.004\\
  HD 222155 & 7.9 & 0.1 & 5834 & 17 & 2.94 & 0.04 & 1.05 & 0.01 & 4.0 & 0.01 & 1.7 & 0.1\\
  HAT-P-6 & 6.3 & 0.2 & 5720 & 54 & 3.9 & 0.1 & 1.13 & 0.02 & 3.85 & 0.03 & 2.1 & 0.1\\
  kappa And & 0.2 & 0.2 & 10942 & 19 & 72.8 & 0.3 & 2.45 & 0.01 & 4.067 & 0.003 & 2.38 & 0.01\\
  WASP-5 & 0.8 & 0.3 & 5819 & 6 & 0.95 & 0.002 & 1.055 & 0.004 & 4.489 & 0.004 & 0.961 & 0.003\\
  HD 224693 & 3.7 & 0.5 & 5960 & 70 & 4.06 & 0.03 & 1.37 & 0.03 & 4.01 & 0.03 & 1.9 & 0.1\\
\end{longtable}
\end{longtab}
}


%
%
%
%
%
%
%
%
%


\begin{thebibliography}{99}
 \bibitem[Baraffe et al. (2005)]{baraffe05} Baraffe I., Chabrier G., Barman T. S., Selsis F., Allard F., Hauschildt P. H. 2005, A\&A, 436, L47
 
 \bibitem[Barker \& Ogilvie, 2009]{barker09} Barker A.J. \& Ogilvie G.I., 2009, MNRAS, 395, 2268-2287. doi:10.1111/j.1365-2966.2009.14694.x
 
 \bibitem[Barnes (2010)]{barnes10} Barnes S.A., 2010, ApJ, 722, 222
 
 \bibitem[Barnes \& Kim (2010)]{barneskim10} Barnes S.A., Kim Y.-C., 2010, ApJ, 721, 675
 
 \bibitem[Bressan et al. (2012)]{bressan12} Bressan A., Marigo P., Girardi L., Salasnich B., Dal Cero C., Rubele S., Nanni A. 2012, MNRAS 427, 127-145
 
 \bibitem[Broeg et al. (2013)]{broeg13} Broeg C., Fortier A., Ehrenreich D., Alibert Y., Baumjohann W., Benz W., Deleuil M., Gillon M., Ivanov A., Liseau R., Meyer M., Oloffson G., Pagano I., Piotto G., Pollacco D., Queloz D., Ragazzoni R., Renotte E., Steller M., Thomas N., the CHEOPS team 2013. arXiv:1305.2270v1
 
 \bibitem[Brown (2014)]{brown14} Brown D.J.A., 2014. arXiv:1406.4402v1
 
 \bibitem[Burgers (1969)]{burgers69} Burgers J.M., 1969, Flow equations for composite gases, Academic Press
 
 \bibitem[Caffau et al. (2008)]{caffau08} Caffau E., Steffen M., Sbordone L., Ludwig H.G., Bonifacio P., 2007, A\&A 473, L9
 
 \bibitem[Chaboyer et al. (2001)]{chaboyer01} Chaboyer B., Fenton W.H., Nelan J.E., Patnaude D.J., Simon F.E. 2001, ApJ, 562:521-527
 
 \bibitem[Chapman \& Cowling (1970)]{chapman70} Chapman S., Cowling T.G., 1970, The mathematical theory of non-uniform gases, Cambridge University Press
 
 \bibitem[da Silva et al. (2006)]{dasilva06} da Silva L., Girardi L., Pasquini L., Setiawan J., von der L\"{u}he O., de Medeiros J. R., Hatzes A., D\"{o}llinger M. P., Weiss A. 2006, A\&A 458, 609-623. doi:10.1051/0004-6361:20065105
 
 \bibitem[Demarque et al. (2004)]{demarque04} Demarque P., Woo J.-H., Kim Y.-C., Yi S.K., 2004, ApJS, 155, 667
 
 \bibitem[Denissenkov (2010)]{denissenkov10} Denissenkov P.A. 2010, ApJ, 719,28:44
 
 \bibitem[Fortney et al. (2007)]{fortney07} Fortney J. J., Marley M. S., Barnes J. W. 2007, ApJ, 659,1661:1672. doi:10.1086/512120
 
 \bibitem[Guenther et al. (1992)]{guenther92} Guenther D.B., Demarque P., Kim Y.C., Pinsonneault M.H., 1992, ApJ, 387, 372
 
 \bibitem[Girardi et al. (2008)]{girardi08} Girardi L., Dalcanton J., Williams B., de Jong R., Gallart C., Monelli M., Groenewegen M.A.T., Holtzman J.A., Olsen K.A.G., Seth A.C., Weisz D.R., 2008, Publications of the Astronomical Society of the Pacific, 120:583-591
 
 \bibitem[Haywood et al. (2013)]{haywood13} Haywood M., Di Matteo P., Lehnert M. D., Katz D., G\'omez A., 2013, A\&A 560, A109. doi:10.1051/0004-6361/201321397
 
 \bibitem[Hubbard et al. (2007)]{hubbard07} Hubbard W. B., Hattori M. F., Burrows A., Hubeny I., Sudarsky D. 2007, Icarus, 187, 358
 
 \bibitem[Hut (1980)]{hut80} Hut P., 1980, A\&A, 92, 167-170
 
 \bibitem[Hut (1981)]{hut81} Hut P., 1981, A\&A, 99, 126-140
 
 \bibitem[J\o rgensen \& Lindegren (2005)]{jorgensen05} J\o rgensen B. R., Lindegren L. 2005, A\&A 436, 127-143. doi:10.1051/0004-6361:20042185
 
 \bibitem[Kroupa et al. (1993)]{kroupa93} Kroupa P., Tout C.A., Gilmore G. 1993, MNRAS, 262, 545
 
 \bibitem[Laughlin \& Chambers (2001)]{laughlin01} Laughlin G. \& Chambers J.E., 2001, ApJ, 551, L109-L113
 
 \bibitem[Liu et al. (2014)]{liu14} Liu K., Bi S.L., Li T.D., Liu Z.E., Tian Z.J., Ge Z.S. arXiv:1406.4402v1
 
 \bibitem[Mamajek \& Hillenbrand (2008)]{mamajek08} Mamajek E.E. \& Hillenbrand L.A. 2008, ApJ, 687:1264-1293
 
 \bibitem[Michaud et al. (1984)]{michaud84} Michaud G., Fontaine G., Baudet G., 1984, ApJ, 282, 206
 
 \bibitem[P\"atzold et al., 2004]{paetzold04} P\"atzold M., Carone L., Rauer H., 2004, A\&A 427, 1075-1080
 
 \bibitem[Pont \& Eyer (2004)]{pont04} Pont F. \& Eyer L. 2004, MNRAS, 351, 487
 
 \bibitem[Poppenhaeger \& Wolk (2014)]{poppenhaeger14} Poppenhaeger K., Wolk S.J. 2014, A\&A 565, L1. doi:10.1051/0004-6361/201423454
 
 \bibitem[Rouan et al. (2012)]{rouan12} Rouan D., Parviainen H., Moutou C., Deleuil M., Fridlund M., Ofir A., Havel M., Aigrain S., Alonso R., Auvergne M., et al., 2012, A\&A 537, A54
 
 \bibitem[Saffe et al. (2005)]{saffe05} Saffe C., G\'omez M., Chavero C., 2005, A\&A 443, 609-626. doi:10.1051/0004-6361:20053452
 
 \bibitem[Seager et al. (2007)]{seager07} Seager S., Kuchner M., Hier-Majumder C. A., Militzer B. 2007, ApJ 669:1279-1297
 
 \bibitem[Soderblom (2010)]{soderblom10} Soderblom D. R., Annu. Rev. Astro. Astrophys., 2010.48:581-629
 
 \bibitem[Sozzetti et al. (2007)]{sozzetti07} Sozzetti A., Torres G., Charbonneau D., Latham D.W., Holman M.J., Winn J.N., Laird J.B., O'Donovan F.T., 2007, ApJ, 664:1190-1198 
 
 \bibitem[Straniero et al. (1992)]{straniero92} Straniero O., Chieffi A., Salaris M. 1992, Mem. S.A.It. vol. 63, n. 2, pag. 315-320
 
 \bibitem[Takeda et al. (2007)]{takeda07} Takeda G., Ford E.B., Sills A., Rasio F.A., Fischer D.A., Valenti J.A 2007, ApJ Supplement Series 168:297-318
 
 \bibitem[Torres (2010)]{torres10} Torres G. 2010, The Astronomical Journal, 140:1158-1162. doi:10.1088/0004-6256/140/5/1158
 
 \bibitem[Winn (2011)]{winn11} Winn J. N. 20 Mar 2011, arXiv:1001.2010v4
 
 \bibitem[Wright et al. (2011)]{wright11} Wright J. T., Fakhouri O., Marcy G. W., Han E., Feng Y., Johnson J. A., Howard A. W., Fischer D. A., Valenti J. A., Anderson J., Piskunov N. 11 Feb 2011, arXiv:1012.5676v3
 
 \bibitem[Wright \& Gaudi (2012)]{wright12} Wright J. T., Gaudi B. S. 11 Oct 2012, arXiv:1210.2471v2
 
 \bibitem[Yang et al. (2013)]{yang13} Yang W., Bi S., Meng X., Liu Z., 2013, ApJ 776:112. doi:10.1088/0004-637X/776/2/112 
\end{thebibliography}
\end{document}